\documentclass[showpacs,aps,prd,reprint,superscriptaddress,nofootinbib,longbibliography,preprintnumbers]{revtex4-2}

\usepackage[colorlinks=true, pdfstartview=FitV, linkcolor=magenta,citecolor=blue, urlcolor=magenta,
bookmarks=true, bookmarksnumbered=true, breaklinks]{hyperref}
\usepackage[dvipdfmx]{graphicx}

\usepackage{here}
\usepackage{url}

\usepackage{amsmath,amssymb,bm,color,longtable,mathrsfs,amsfonts,slashed}

\newcommand{\Slash}[1]{{\ooalign{\hfil/\hfil\crcr$#1$}}}

\renewcommand\b{\beta}

\renewcommand\k{\kappa}
\renewcommand\l{\lambda}

\renewcommand\c{\chi}

\newcommand\e{\epsilon}
\newcommand\g{\gamma}

\newcommand\m{\mu}

\newcommand\p{\pi}

\newcommand\s{\sigma}

\newcommand\w{\eta}

\renewcommand\P{\Pi}

\renewcommand\O{\Omega}

\newcommand\D{\Delta}

\newcommand\F{\Phi}

\newcommand\fl{\mathcal{L}}
\newcommand\mtr{\mathrm{Tr}}

\newcommand\mln{\mathrm{ln}}

\newcommand\me{\mathrm{e}}
\newcommand\mi{\mathrm{i}}

\begin{document}
\begin{flushright}
\end{flushright}

\title{Meson mass spectrum in isospin QCD medium from the $N_f=2+1$ quark-meson model}

\author{Yu-Han Gao}
\email{yhgao24@m.fudan.edu.cn}
\affiliation{Physics Department and Center for Particle Physics and Field Theory, Fudan University, Shanghai 200438, China}
\author{Xu-Guang Huang}
\email{huangxuguang@fudan.edu.cn}
\affiliation{Physics Department and Center for Particle Physics and Field Theory, Fudan University, Shanghai 200438, China}
\affiliation{Key Laboratory of Nuclear Physics and Ion-beam Application (MOE), Fudan University, Shanghai 200433, China}
\affiliation{Shanghai Research Center for Theoretical Nuclear Physics, National Natural Science Foundation of China and Fudan University, Shanghai 200438, China}
\author{Daiki Suenaga}
\email{suenaga.daiki.j1@f.mail.nagoya-u.ac.jp}
\affiliation{Kobayashi-Maskawa Institute for the Origin of Particles and the Universe, Nagoya University, Nagoya 464-8602, Japan}
\affiliation{Research Center for Nuclear Physics, Osaka University, Ibaraki 567-0048, Japan}

\date{\today}

\begin{abstract}
We study phase structures, meson mass spectra, and sound velocities at finite temperature and density in QCD with isospin chemical potential ($\mu_I$). We employ the quark-meson model with $N_f=2+1$, incorporating the Kobayashi-Maskawa-'t Hooft type coupling to capture dynamical effects of the $U(1)_A$ axial anomaly. Within the mean-field approximation at quark one-loop, we analyze the onset of pion condensation, which manifests as a second-order phase transition at low temperatures and may exhibit a first-order behavior at higher temperatures within this approximation. We examine the corresponding mass spectra of scalar and pseudoscalar singlet-octet mesons, in which the $\pi_+$ mass is exactly massless in the superfluid phase due to its Nambu-Goldstone boson nature. The neutral pion mass is, meanwhile, found to exhibit a strictly linear growth with $\mu_I$ in the pion condensed phase. We also investigate the isothermal squared sound velocity and identify characteristic structures associated with the phase transitions. Furthermore, we highlight how enhanced $U(1)_A$ anomaly effects facilitate the pion condensate to generate a less-pronounced sound velocity peak in cold medium. Our findings are expected to provide future lattice simulations with useful information on meson mass spectra from symmetry aspects.
\end{abstract}


\maketitle

\section{Introduction}
\label{sec:Introduction}

Exploration of quantum chromodynamics (QCD) in hot and/or dense medium has been attracting much attention, since noteworthy transition such as color deconfinement and chiral-symmetry restoration are expected in such extreme conditions~\cite{Fukushima:2010bq}. While heavy-ion collision (HIC) experiments have shaded light on hot QCD properties in the past two decades~\cite{Busza:2018rrf}, in order to gain insights into cold and dense medium we need to rely on neutron star (NS) observations which are less controllable. Therefore, dense QCD properties remain wrapped in mystery compared to hot medium.

One of the promising tools to explore QCD properties is the {\it lattice QCD simulation}, which is based on a first-principle numerical simulation on the discretized Euclidean space. The lattice QCD is straightforwardly applicable at finite temperature as well as in the vacuum, unless a (baryon) chemical potential is added. At finite chemical potential, however, the cumbersome {\it sign problem} prevents us from making use of the Monte-Carlo simulations, and hence it is not easy to operate the powerful lattice QCD there~\cite{sign-problem1,sign-problem2,sign-problem3,sign-problem4,Bazavov:2011nk}. This problem is another reason why our understandings of cold and dense QCD are limited.

Although the sign problem appears in lattice simulations with finite baryon chemical potential, it cancels in QCD with an isospin chemical potential ($\mu_I$) as long as the corresponding isospin symmetry is imposed. For this reason, QCD with finite $\mu_I$, i.e., QCD$_I$, offers a useful testing ground to examine hadron and quark-gluon dynamics in cold medium. Thus far, many efforts from lattice simulations in QCD$_I$ medium have been made so as to reveal dense properties~\cite{Kogut:2002zg,Kogut:2004zg,Detmold:2012wc,Brandt:2017oyy,Brandt:2022hwy}, with which theoretical studies such as the chiral perturbation theory (ChPT)~\cite{PhysRevLett.86.592,Son:2000by,Splittorff:2000mm,Andersen:2023ofv,Adhikari:2020ufo}, Nambu-Jona-Lasinio (NJL) model~\cite{He:2005nk,Xia:2013caa,Liu:2021gsi,Zhang:2025stm}, Polyakov-loop-based model~\cite{Bratovic:2012qs,Stiele:2013pma,Chiba:2024cny,Adhikari:2018cea}, Dyson-Schwinger equation (DSE)~\cite{Xin:2014ela,Fischer:2014ata,Xu:2021dki}, and functional renormalization group (FRG)~\cite{Kamikado:2012bt,Fu:2019hdw,Gao:2020qsj,Svanes:2010we} have been done. It should be noted that QCD-like system such as two-color QCD ($N_c=2$) with baryon chemical potential is also regarded as a promising testing ground, thanks to its sign-problem-free nature
~\cite{Boz:2019enj,Buividovich:2020dks,Astrakhantsev:2020tdl,Iida:2024irv}.

Isospin matter is created by hadrons carrying isospin charges. The lightest ones are charged pions, which are bosons, and thus QCD$_I$ matter will form a Bose-Einstein-condensation (BEC) phase of charged pions when $\mu_I$ reaches a critical value depending on temperature~\cite{PhysRevLett.86.592,Son:2000by}. This BEC phase does not break any color symmetries, so it is referred to as the {\it pion superfluid phase}.

In the mean-field approximation, the phase transition to the pion superfluid phase is predicted to be of second-order at lower temperatures. At adequately higher temperatures, some model analyses predict that the second-order transition can turn into the first-order one~\cite{Ferreira:2025zeu,Kamikado:2012bt}, while examinations incorporating fluctuations claim that the transition is always of second-order~\cite{Zhang:2006dn,Xu:2021dki}, and lattice simulations may support the latter~\cite{Brandt:2022hwy}.
 
In this study, we investigate meson mass spectra in QCD$_I$ as a complementary source of information on the nature of the phase transition. In hot and dense QCD$_I$ matter, quark degrees of freedom are expected to become increasingly relevant as the system approaches the deconfined regime. For this reason, we employ a model incorporating both mesons and quarks with the linear realization of chiral symmetry, i.e., the so-called quark-meson model~\cite{PhysRevD.97.076005,Schaefer_2005,Ueda_2013,Schaefer_2009}. 
Here, it should be emphasized that meson masses serve not only as fundamental observables but also as highly sensitive probes to the symmetry properties in medium~\cite{PhysRevLett.86.592,PhysRevLett.111.021601,PhysRevLett.110.011602}. Hence, we adopt the $N_f=2+1$ version of the quark-meson model where the $U(1)$ axial anomaly induced flavor-mixing structures is incorporated by the Kobayashi-Maskawa-'t Hooft (KMT) type interaction~\cite{tHooft1976,Kobayashi:1970ji,Kunihiro:2009ds}, in order to gain deeper insights into the anomaly effects as well.

The strength of the $U(1)_A$ anomaly has a crucial influence on determining QCD phase structure~\cite{Lenaghan:2000ey,Fukushima:2001hr,Schaefer:2008hk,Costa:2008dp,Ruivo:2012xt}. In extremely high temperature and/or density, the anomaly effects on hadrons, which are fundamentally driven by the instantons, are expected to be suppressed due to the Debye screening of electric gluons. Meanwhile, FRG analyses imply that the anomaly effects are once enhanced in the intermediate temperature and/or density due to hadronic fluctuations~\cite{Fejos:2015xca,Fejos:2021yod}. Hence, we also present results by changing the strength of the KMT coupling so as to take a look at responses to variation of the anomaly effects.

Although meson mass spectra contain much information on the ground-state structure of matter from symmetry aspects, it is also invaluable to examine the equation of state (EoS) which plays a basic role in characterizing dense medium~\cite{Annala:2019puf,Herbst:2013ail}. Astrophysically the ``stiffness'' of EoS provides significant information on the mass-radius measurements and gravitational-wave observations of NSs~\cite{Antoniadis:2013pzd,Fonseca:2016tux,NANOGrav:2017wvv,LIGOScientific:2018cki,NANOGrav:2019jur,Fonseca:2021wxt,Miller_2019,Miller_2021}. An alternative observable that is capable of intuitively representing soft or stiff EoS is the sound velocity. On the lattice, the existence of a peak structure in the isothermal squared sound velocity $c_T^2$ has been revealed~\cite{Iida:2022hyy,Brandt:2022hwy,Abbott:2023coj}, before reaching the conformal limit $c_s^2=1/3$ in extremely dense regime. We study the sound velocity to confirm the peak structure and see what happens in the presence of the possible first-order phase transition, within the quark-meson model. 

In this work, employing the mean-field approximation with quark one-loop corrections, we find that the pion-superfluid transition, which is of second order at low temperatures, may exhibit first-order behavior at higher temperatures. We further investigate the complete meson mass spectrum across the phase boundary and find that the meson masses, particularly the lowest $\pi_+$ mass, show pronounced discontinuous changes in the corresponding region. We also demonstrate that the $\pi^+$ mode becomes the exact Nambu-Goldstone (NG) boson associated with the spontaneous breaking of the residual $U(1)_{\l_3}$ symmetry in the superfluid phase, while the massive pion exhibits a robust linear dependence on $\mu_I$, at any temperatures.

Besides, we investigate the EoS of isospin matter and sound velocity. It is found that within the present quark-meson model, the EoS experiences a transient stiffening upon entering the superfluid phase. 

Furthermore, by changing the strength of KMT coupling, we find that stronger anomaly effects lower the chiral crossover temperature while enhances pion condensate in cold and dense medium. It is also found that the stronger anomaly leads to less-pronounced peak structures in the sound velocity.

The present paper is organized as follows. In Sec.~\ref{sec:model} we introduce the quark-meson model with $N_f=2+1$ and explain general properties in the pion superfluid phase. Our procedure to fix the model parameters is also illustrated. Then, in Sec.~\ref{sec:phase} we present the phase structure in hot and dense QCD$_I$. Meson mass spectra are shown in Sec.~\ref{sec:meson}, while the EoS and sound velocities are analyzed in Sec.~\ref{sec:EoS}. The role of the axial anomaly is investigated in Sec.~\ref{sec:anomaly}. Finally, Sec.~\ref{sec:summary} summarizes our results and outlines possible future developments.

\section{Model}
\label{sec:model}

\subsection{Quark-meson model Lagrangian}
\label{sec:QMLagrangian}

In order to describe hadronic and quark degrees of freedom in a unified way in broad range of QCD$_I$ medium, we adopt the $N_f=2+1$ quark-meson model~\cite{Tetradis:2003qa,Schaefer:2004en,Stiele:2013pma,Ayala:2026pzu}, whose Lagrangian is given by
\begin{equation}
\fl_{\rm QM}=\fl_q+\fl_M + \fl_{\rm ex.}\ . \label{LQMM}
\end{equation}
In this Lagrangian, the first piece
\begin{eqnarray}
{\cal L}_q = \bar{q}i\Slash{D}q -g(\bar{q}_LM q_R + \bar{q}_RM^\dagger q_L) \label{Lquark}
\end{eqnarray}
stands for the quark parts coupled to the mesons, where $q=(u,d,s)^T$ is a quark triplet ($q_{R/L} = \frac{1\pm\gamma_5}{2}q$) and the covariant derivative~\cite{Loewe:2005df,Son:2000by} is defined by $D_\mu q = (\partial_\mu-i\mu_I\lambda_3\delta_{\mu0})q$ with the isospin chemical potential $\mu_I$. $\lambda_a$ ($a=0$--$8$) are the Gell-Mann matrices ($\lambda_0=\sqrt{\frac{2}{3}}{\bm 1}$). The meson matrix $M$ takes the form of
\begin{eqnarray}
M = \sum_{a=0}^8(\sigma_a+i\pi_a)\lambda_a\ , 
\end{eqnarray}
where $\s_a$ and $\p_a$ represent the scalar and pseudoscalar meson nonets, respectively. Under $SU(3)_L\times SU(3)_R$ chiral transformation, the left-handed and right-handed quark fields transform as $q_R \to U_Rq_R$ and $q_L\to U_L q_L$, respectively, with $U_{R/L}\in SU(3)_{R/L}$, while the meson nonet $M \to U_LM U_R^\dagger$. Hence the Yukawa interaction in Eq.~(\ref{Lquark}) preserves $SU(3)_L\times SU(3)_R$ chiral symmetry~\cite{GellMann1960TheAV}.

In Eq.~(\ref{LQMM}), the second piece ${\cal L}_M$ represents meson kinetic terms and self interactions
\begin{eqnarray}
\fl_{M} &=&\frac{1}{4}{\rm tr}_f[(D_\mu M)^\dagger D^\mu M]-\frac{m^2_M}{4}{\rm tr}_f[M^\dagger M] \nonumber\\
&& +\frac{\l_1}{48}({\rm tr}_f[M^\dagger M])^2  +\frac{\l_2}{48}{\rm tr}_f[(M^\dagger M)^2] \nonumber\\
&& -\frac{K}{2}(\det M+\det M^\dagger) \ , \label{LMeson}
\end{eqnarray}
with $D_\mu M = \partial_\mu M-i\mu_I[\lambda_3,M]\delta_{\mu0}$,\footnote{The $\mu_I$ for mesons $M$ can be systematically incorporated by gauging with respect to $SU(2)_V$ and replacing the corresponding gauge field $V_\mu$ as $V_\mu \to \mu_I\lambda_3\delta_{\mu0}$.} where the symbol ``tr$_f$'' stands for the trace operator acting on the flavor space. In this equation, we have introduced the interactions up to ${\cal O}(M^4)$ which are invariant under $SU(3)_L\times SU(3)_R$ chiral and $U(1)_B$ transformations. The last $K$ term is the KMT-determinant type interaction which violates $U(1)$ axial symmetry. This coupling is incorporated to take into account the $U(1)$ axial anomaly effects~\cite{10.1143/PTP.45.1955,PhysRevD.14.3432}.

The last piece in Eq.~(\ref{LQMM}) corresponds to an external source term
\begin{eqnarray}
{\cal L}_{\rm ex.} &=& \frac{c}{2}{\rm tr}[{\cal M}_q^\dagger M+M^\dagger{\cal M}_q] \nonumber\\
&=& h_l\sigma_l + \frac{h_s}{2}\sigma_s
\end{eqnarray}
which is responsible for finite current-quark mass effects, with ${\cal M}_q = {\rm diag}(m_l,m_l,m_s)$ being a current-quark mass matrix under $SU(2)$ isospin symmetry. In the second line, we have defined $h_l=2cm_l$ and $h_s=2cm_s$ for convenience, and
\begin{eqnarray}
\sigma_l = \sqrt{\frac{2}{3}}\sigma_0 + \frac{1}{\sqrt{3}}\sigma_8\ , \ \ \sigma_s = \frac{1}{\sqrt{3}}\sigma_0 - \sqrt{\frac{2}{3}}\sigma_8\ ,
\end{eqnarray}
for which their quark contents are manifestly understood. In the following analysis, we will adopt the same basis for $\pi_0$ and $\pi_8$:
\begin{eqnarray}
\pi_l = \sqrt{\frac{2}{3}}\pi_0 + \frac{1}{\sqrt{3}}\pi_8\ , \ \ \pi_s = \frac{1}{\sqrt{3}}\pi_0 - \sqrt{\frac{2}{3}}\pi_8\ .
\end{eqnarray}

We note that $SU(2)$ isospin symmetry is broken due to the nonzero $\mu_I$, but there remains a residue symmetry $U(1)_{\l_3}$ generated by a rotation $U_{\lambda_3}(\theta_3) = {\rm e}^{-i\theta_3 \lambda_3}$. Therefore, in the presence of $\mu_I$, the remaining symmetries are $U(1)_B\times U(1)_S\times U(1)_{\lambda_3}$, where the former two correspond to the baryon-number and strange-number symmetry.

The isospin chemical potential provides an energy gain for excitations corresponding to their isospin charges; the single-particle excitation energy of a particle $\alpha$ at rest is determined by $E_\alpha = m_\alpha -\mu_IQ_I^\alpha$, where $m_\alpha$ and $Q_I^\alpha$ stand for the mass and isospin charge of $\alpha$. Hence, the positively charged pion becomes the lightest excitation at finite $\mu_I$ and dominates over the low-energy dynamics. In particular, since $Q_I^{\pi_+} =+2$, when $\mu_I$ reaches $\mathring{M}_{\pi}/2$ with $\mathring{M}_\pi$ being the vacuum pion mass,\footnote{Throughout this paper, we use a symbol ``$\mathring{X}$'' to refer to the vacuum value of $X$.}  the energy gain associated with the isospin charge completely compensates the energy cost of creating a $\pi_+$, in the absence of thermal fluctuations. Consequently, the normal vacuum becomes unstable against creating $\pi_+$ for $\mu_I\geq \mathring{M}_\pi/2$ such that macroscopic occupation of $\pi_+$, i.e., the BEC of $\pi_+$, manifests itself. This peculiar phase with pion condensates is referred to as the pion superfluid phase~\cite{PhysRevLett.86.592,Son:2000by,Kogut:2002tm,He:2005nk,PhysRevD.82.016005,Nishihara:2013nem}. In the present study, we restrict ourselves in a range of $\mu_I$ where only pion condensations emerge, although there remains a possibility of condensates created by other hadrons carrying nonzero isospin charge~\cite{Aharony_2008,SCHAFER200167,Brauner:2016lkh}. Thus we take into account vacuum expectation values (VEVs) of $\sigma_l$, $\sigma_s$ and $\pi_1$ corresponding to the chiral and pion condensations with an appropriate phase choice, which leads to
\begin{equation}
\langle M \rangle=
\begin{bmatrix}
\left \langle \s_l \right \rangle   & \mi\left \langle  \p_1 \right \rangle  & 0\\
 \mi\left \langle  \p_1 \right \rangle & \left \langle \s_l \right \rangle & 0\\
 0 & 0 & \sqrt{2}\left \langle \s_s \right \rangle
\end{bmatrix}
\ .
\label{MeanField}
\end{equation}
We note that the critical point of the superfluidity deviates from $\mu_I=\mathring{M}_\pi/2$ at finite temperatures due to thermal fluctuations.

\subsection{Discrete symmetry in the pion superfluid phase}
\label{sec:Symmetry}

While in the normal phase the system possesses $U(1)_B\times U(1)_S \times U(1)_{\l_3}$ symmetries~\cite{Lenaghan:2000ey}, in the pion superfluid phase they are broken down into $U(1)_B\times U(1)_S$ due to nonzero $\langle\pi_1\rangle$. Since $\pi_1$ is a pseudoscalar, parity is no longer a good quantum number in the latter phase, which results in various mixings among scalar and pseudoscalar mesons. One would infer that all the $\sigma_a$'s and $\pi_a$'s could mix as long as strange number is counted, but this is not the case. In this subsection we explain this property~\cite{HAO2007275}.

The operator $\pi_1\sim \bar{q}i\gamma_5\lambda_1 q$ is odd under a discrete transformation generated by $U_{\lambda_3}\left(\frac{\pi}{2}\right) = {\rm e}^{-i\frac{\pi}{2}\lambda_3}$:
\begin{eqnarray}
 \bar{q}i\gamma_5 \lambda_1 q \to \bar{q}U^\dagger_{\lambda_3}\left(\frac{\pi}{2}\right)i\gamma_5 \lambda_1 U_{\lambda_3}\left(\frac{\pi}{2}\right)q = - \bar{q}i\gamma_5 \lambda_1 q \ ,
\end{eqnarray}
owing to $\{\lambda_1,\lambda_3\} = 0$ and $U_{\lambda_3}(\pi) = -1$. Thus, since $\pi_1$ is a pseudoscalar, one can see that $\pi_1$ is invariant under a discrete symmetry generated by
\begin{eqnarray}
\hat{U}_{\lambda_3}\left(\frac{\pi}{2}\right) \equiv \hat{P} U_{\lambda_3}\left(\frac{\pi}{2}\right) = \hat{P}{\rm e}^{-i\frac{\pi}{2}\lambda_3} \label{UHat}
\end{eqnarray}
with $\hat{P}$ representing the parity operator, although the continuous $U(1)_{\l_3}$ is spontaneously broken. In a similar way one can easily confirm that $\pi_1$ is invariant under
\begin{eqnarray}
\hat{V}_{\lambda_1}\left(\frac{\pi}{2}\right) \equiv \hat{C}{\rm e}^{-i\frac{\pi}{2}\lambda_1}\ , \label{VHat}
\end{eqnarray}
with a charge-conjugation operator $\hat{C}$. Therefore, the ground state in the pion superfluid phase preserves discrete symmetries generated by Eqs.~(\ref{UHat}) and~(\ref{VHat}).\footnote{Since $\sigma_l\sim \frac{1}{\sqrt{2}}(\bar{u}u+\bar{d}d)$ and $\sigma_s\sim \bar{s}s$ are even under parity or charge conjugation, it is obvious that  $\sigma_l$ and $\sigma_s$ are invariant under the discrete transformations by Eqs.~(\ref{UHat}) and~(\ref{VHat}).}

The above discrete symmetries divide $18$ mesons into six groups as summarized in Table~\ref{tab:meson}. In this table the second and third columns exhibit how the mesons are transformed by Eqs.~(\ref{UHat}) and~(\ref{VHat}), respectively. The mass matrix of mesons in the superfluid phase becomes block-diagonal in this basis, for which the mixing occurs within each sector.

\begin{table}[h]
\centering
\begin{ruledtabular}
\begin{tabular}{ccc}
mesons & $\hat{U}_{\lambda_3}(\frac{\p}{2})$  & $\hat{V}_{\lambda_1}(\frac{\p}{2})$  \\
\hline
$\p_3$ & $-\p_3$ & $-\p_3$   \\
\hline
$\s_3$ & $\s_3$ & $-\s_3$  \\
\hline
$(\p_1,\p_2,\s_l,\s_s)$ & $(\p_1,\p_2,\s_l,\s_s)$ & $(\p_1,\p_2,\s_l,\s_s)$\\
\hline
$(\p_l,\p_s,\s_1,\s_2)$ & $-(\p_l,\p_s,\s_1,\s_2)$ & $(\p_l,\p_s,\s_1,\s_2)$ \\
\hline
$(\p_4,\p_5,\s_6,\s_7)$ & $(-\p_5,\p_4,-\s_7,\s_6)$ & undefined \\
\hline
$(\p_6,\p_7,\s_4,\s_5)$ & $(\p_7,-\p_6,\s_5,-\s_4)$ & undefined \\
\end{tabular}
\end{ruledtabular}
\caption{
Six groups obtained from the discrete transformations generated by Eqs.~(\ref{UHat}) and~(\ref{VHat}).}
\label{tab:meson}
\end{table}

\subsection{Quark propagator in the pion superfluid phase}
\label{sec:QMMedium}

In this subsection, based on the quark-meson Lagrangian introduced in Sec.~\ref{sec:QMLagrangian} we present the thermodynamic potential at quark one-loop and derive the quark propagator which will play a central role in evaluating meson masses in medium.

The thermodynamic potential per unit volume at quark one-loop level (mean field approximation) within our quark-meson model~(\ref{LQMM}) is derived by means of, e.g., the imaginary-time formalism~\cite{Kapusta:2006pm}, which yields ($N_c=3$)
\begin{eqnarray}
\O &=& -N_c T\sum_n\int_{\vec{p}} {\rm tr}_{f}\ln[S^{-1}(p)] + V_{\rm MF} \nonumber\\
&=& -4N_c\sum_{\xi={\rm p,a}}\int_{\vec{p}}\Bigg[\left(\frac{\e_{\xi}}{2}+\frac{E_s}{2}\right) \nonumber\\
&& + T\mln\left(1+\me^{-\e_\xi/T} \right)+T\mln\left(1+\me^{- E_s/T} \right) \Bigg] +V_{\rm MF}\ , \nonumber\\
\label{eq:thermal potential}
\end{eqnarray}
where the mesonic potential at mean-field level reads
\begin{eqnarray}
V_{\rm MF} &=&-\frac{2\m_I^2}{g^2}\D^2 -\frac{m^2_M}{4g^2}\left[2(M_l^2+\D^2)+M_s^2\right] \nonumber\\
&&+\frac{\l_1}{48g^4}(2M_l^2+2\D^2+M_s^2)^2 \nonumber\\
&&+\frac{\l_2}{48g^4}\left[2(M_l^2+\D^2)^2 + M_s^4\right] \nonumber\\
&&-\frac{h_l}{g}M_l-\frac{h_s}{2g}M_s-\frac{K}{g^3}(M_l^2+\D^2)M_s \ ,
\label{MFPotential}
\end{eqnarray}
with the dynamical quark masses and gap defined by
\begin{equation}
M_l=g\langle \s_l \rangle,\quad M_s=\sqrt{2}g\langle \s_s \rangle,\quad \D=g\langle \p_1 \rangle\ . \label{Gaps}
\end{equation}
In the first line in Eq.~(\ref{eq:thermal potential}), $\int_{\vec{p}} = \int d^3p/(2\pi)^3$. $S_q^{-1}(p)$ is the inverse quark propagator read off from the quadratic terms in Eq.~(\ref{Lquark}) with which the mean fields~(\ref{MeanField}) are inserted. That is,
\begin{eqnarray}
&& S_q^{-1}(p) \nonumber\\
&=&\begin{pmatrix}
\Slash{p}+\gamma^0\mu_I-M_l & -\mi\gamma^5\Delta & 0\\
 -\mi\gamma^5\Delta & \Slash{p}-\gamma^0\mu_I-M_l & 0\\
 0 & 0 & \Slash{p}-M_s
\end{pmatrix}\ ,
\label{SInverse}
\end{eqnarray}
where $p_0=i\omega_n$ with $\omega_n=(2n+1)\pi T$ ($n\in \mathbb{Z}$) being the fermionic Matsubara frequency. In the second equality in Eq.~(\ref{eq:thermal potential}), we have defined single-particle energies of the quasiparticle upon the pion superfluidity and of $s$ quark by 
\begin{eqnarray}
\e_\xi(\vec{p}) &=& \sqrt{(E_l(\vec{p}) - \eta_\xi\m_I)^2+\D^2}\ , \nonumber\\
E_s(\vec{p}) &=& \sqrt{\vec{p}^2+M_s^2}\ ,
\end{eqnarray}
with $E_l(\vec{p}) = \sqrt{\vec{p}^2+M_l^2}$ and $\eta_{\rm p} = +1$ while $\eta_{\rm a} = -1$ The subscripts $\rm p$ and $\rm a$ correspond to the particle and antiparticle contributions of the light quarks, respectively.

The quark propagator matrix is obtained by inverting Eq.~(\ref{SInverse}), which results in~\cite{He:2005nk,Suenaga:2021bjz}
\begin{equation}
S_q(p)=\begin{pmatrix}
 S_{uu} & S_{ud} & 0\\
 S_{du} & S_{dd} & 0\\
 0 & 0 & S_{ss}
\end{pmatrix}\ ,
\label{SMatix}
\end{equation}
where
\begin{eqnarray}
S_{uu} &=& \sum_{\xi={\rm p,a}}\left(\frac{|u_\xi(\vec{p})|^2}{p_0-\eta_\xi\epsilon_\xi(\vec{p})} + \frac{|v_\xi(\vec{p})|^2}{p_0+\eta_\xi\epsilon_\xi(\vec{p})}\right)\Lambda^l_\xi\gamma_0 \ , \nonumber\\
S_{ud} &=& - \sum_{\xi={\rm p,a}}\left(\frac{u^*_\xi(\vec{p})v^*_\xi(\vec{p})}{p_0-\epsilon_\xi(\vec{p})} - \frac{u^*_\xi(\vec{p})v^*_\xi(\vec{p})}{p_0 + \epsilon_\xi(\vec{p})} \right)\Lambda^l_\xi\gamma_5 \ , \nonumber\\
S_{du} &=& \sum_{\xi={\rm p,a}}\left(\frac{u_\xi(\vec{p})v_\xi(\vec{p})}{p_0-\epsilon_\xi(\vec{p})} - \frac{u_\xi(\vec{p})v_\xi(\vec{p})}{p_0 + \epsilon_\xi(\vec{p})} \right)\Lambda^{l,C}_\xi\gamma_5  \ , \nonumber\\
S_{dd} &=& \sum_{\xi={\rm p,a}}\left(\frac{|v_\xi(\vec{p})|^2}{p_0-\eta_\xi\epsilon_\xi(\vec{p})} + \frac{|u_\xi(\vec{p})|^2}{p_0+\eta_\xi\epsilon_\xi(\vec{p})}\right)\Lambda^{l,C}_\xi\gamma_0 \ .  \label{SlightElement}\nonumber\\
\end{eqnarray}
In these expressions, coherence factors are defined in such a way as to satisfy
\begin{eqnarray}
&& |u_{\xi}(\vec{p})|^2 = \frac{1}{2}\left(1+\frac{E_l(\vec{p})-\eta_\xi\mu_I}{\epsilon_\xi(\vec{p})} \right)\ , \nonumber\\
&& |v_\xi(\vec{p})|^2 = \frac{1}{2}\left(1-\frac{E_l(\vec{p})-\eta_\xi\mu_I}{\epsilon_\xi(\vec{p})} \right)\ , 
\end{eqnarray}
and
\begin{eqnarray}
 |u_\xi(\vec{p})|^2 +  |v_\xi(\vec{p})|^2  = 1 \ , \ \ u_\xi(\vec{p})v_\xi(\vec{p}) = -\mi\frac{\Delta}{2\epsilon_\xi(\vec{p})}\ .
\end{eqnarray}
The matrices $\Lambda^l_\xi$ are the positive-energy and negative-energy projection operators ($\vec{\alpha} = \gamma_0\vec{\gamma}$, $\beta=\gamma_0$)
\begin{eqnarray}
\Lambda^l_\xi = \frac{E_l(\vec{p}) + \eta_\xi(\vec{\alpha}\cdot\vec{p} + \beta M_l )}{2E_l(\vec{p})}\ , \ \ \Lambda_{\rm p/a}^{l,C} = \Lambda^l_{\rm a/p}\ . \label{Projection}
 \end{eqnarray}

The component $S_{ss}$ in Eq.~(\ref{SMatix}) is straightforwardly obtained as
\begin{eqnarray}
S_{ss} = \frac{1}{\Slash{p}-M_s} =\sum_{\xi={\rm p,a}}\frac{\Lambda_\xi^s}{p_0-\eta_\xi E_s(\vec{p})}\ , \label{Sstrange}
\end{eqnarray}
with
\begin{eqnarray}
\Lambda^s_\xi = \frac{E_s(\vec{p}) + \eta_\xi(\vec{\alpha}\cdot\vec{p} + \beta M_s )}{2E_s(\vec{p})}\ .
\end{eqnarray}

In principle, the momentum integrals in Eq.~(\ref{eq:thermal potential}) suffer from ultraviolet (UV) divergences owing to vacuum fluctuations at $\mu_I=0$ and $T=0$, which requires a renormalization. Although there are many choices of the renormalization scheme, as long as we do not consider high energy scale, e.g., sufficiently high density or high temperature, all the schemes are expected to lead to qualitatively similar results~\cite{Koch:1997ei}. We therefore will neglect contributions which are explicitly independent of $T$ in the following analysis, using the so-called {\it no-sea approximation} as one of the simplest regularization scheme \cite{Ferreira:2025zeu,Rai:2023vnp} (for a discussion on the relevance of vacuum terms,
see Appendix~\ref{ap:no-sea} and Refs.~\cite{Kamikado:2012bt,Skokov:2010sf,Palhares:2010be,PhysRevC.70.015204,PhysRevD.78.025013,Fraga:2009pi}). It should be noted that within the no-sea approximation, quark one-loop contributions only enter at finite temperature.

The true ground state of an equilibrium system is determined by finding the global minimum of the thermodynamic potential with respect to the condensates. The corresponding gap equations and related discussions are supplemented in Appendix~\ref{ap:GapEquation}.

\subsection{Parameters fixing}
\label{sec:Input}

Our present work aims to investigate meson mass spectra and other quantities in QCD$_I$ medium, and to explore their usefulness for characterizing the phase structure of QCD$_I$ and for future comparisons with lattice QCD. To this end, here we explain how the model parameters are fixed.

Our model involves seven free parameters: $m_M^2$, $\lambda_1$, $\lambda_2$, $K$, $h_l$, $h_s$ and $g$. It should be noted that the mean fields $\langle\sigma_l\rangle$, $\langle\sigma_s\rangle$ and $\langle\pi^1\rangle$ are determined by means of the respective gap equations.
First, we employ the experimentally measured pion and kaon decay constants, $f_\pi$ and $f_K$, as inputs:
\begin{eqnarray}
f_\pi = 0.0924\, {\rm GeV}\ , \ \ f_K = 0.1336\, {\rm GeV}\ .
\end{eqnarray}
Within our quark-meson model, those decay constants are evaluated by relations $f_\pi Z_\pi^{1/2} = \langle\mathring{\sigma}_l\rangle$ and $f_K Z_K^{1/2} = \langle\mathring{\sigma}_l+\sqrt{2}\mathring{\sigma}_s\rangle/2$, with pion and kaon renormalization factors $Z_\pi$ and $Z_K$. In the vacuum the quark one-loop contributions are omitted owing to the no-sea approximation, which results in $Z_\pi=Z_K=1$ (at $T=\mu_I=0$). Hence, simply $f_\pi = \langle\mathring{\sigma}_l\rangle$ and $f_K  = \langle\mathring{\sigma}_l+\sqrt{2}\mathring{\sigma}_s\rangle/2$.

Next, we adopt the observed masses of pion, kaon, $f_0(500)$, and $\eta$ and $\eta'$ mesons as further inputs:
\begin{eqnarray}
&&\mathring{M}_\pi^2= (0.138\, \text{GeV})^2\ , \ \ \mathring{M}^2_{K}=(0.497\, \text{GeV)}^2 \ , \nonumber\\
&& \mathring{M}_{f_0(500)}^2 = \mathring{M}^2_{\s_l}=(0.600\, \text{GeV)}^2\ , \nonumber\\
&& \mathring{M}^2_{\w}+\mathring{M}^2_{\w '} = (0.547\, \text{GeV)}^2+(0.958\, \text{GeV)}^2\ .
\end{eqnarray}
While we have used a combination of $\mathring{M}^2_{\w}+\mathring{M}^2_{\w '}$ as an input~\cite{Zacchi:2015lwa}, the resultant $\eta$ and $\eta'$ masses read $\mathring{M}_\w=0.540\, \text{GeV}$ and $\mathring{M}_{\eta'}=0.962\, \text{GeV}$, which are reasonably close to the experimental values. The mass formulas in the presence of $\Delta$ as well as $M_l$ and $M_s$ are supplemented in Appendix~\ref{ap:meson mass}.

\begin{table}[t]
\begin{center}
  \begin{tabular}{ccccccc}  \hline\hline
$m_M^2$ & $\lambda_1$ & $\lambda_2$ & $K$ & $h_l$ & $h_s$ & $g$ \\ \hline
$-0.118$ & $3.62$ & $141$ & $1.20$ & $1.76\times 10^{-3}$ & $5.41\times 10^{-2}$ & $3.15$ \\
\, [GeV$^{2}$] & & & [GeV] & [GeV$^3$] & [GeV$^3$] & \\ 
\hline \hline
 \end{tabular}
\caption{The determined model parameters.}
\label{tab:Parameter}
\end{center}
\end{table}

Finally, we choose the Yukawa coupling to be $g=3.15$ which results in the dynamical quark masses of $\mathring{M}_l=291\, \text{MeV},\mathring{M}_s=421\, \text{MeV}$ in the vacuum. The determined parameters are summarized in Table~\ref{tab:Parameter}.

\section{phase structure of QCD$_I$}
\label{sec:phase}
\subsection{pion condensate}

As illustrated in Sec.~\ref{sec:QMLagrangian}, the charged pions form BECs when $\mu_I$ exceeds the critical value: $\mu_I^c = \mathring{M}_\pi/2$, to enter the pion superfluid phase, at vanishing temperature. This critical value is also explicitly seen from the gap equation~(\ref{eq:gap equation}); the gap equation for an infinitesimally small $\Delta$ is reduced to
\begin{eqnarray}
\mu_I^2-\left(\frac{\mathring{M}_\pi}{2}\right)^2 -g^2N_c\sum_{\xi={\rm p,a}}\int_{\vec{k}}\frac{f(|E_l-\eta_\xi\mu_I|)}{|E_l-\eta_\xi \mu_I|}= 0\  \label{GapEqDelta0}
\end{eqnarray}
with the Fermi-Dirac distribution function ($\beta=1/T$)
\begin{equation}
f(x)=\frac{1}{\me^{\b x} +1}\ , 
\end{equation} 
which claims that the critical value is given by  $\mu_I^c = \mathring{M}_\pi/2$ as long as $T=0$. This not only ensures the mathematical self-consistency but also naturally satisfies the physical requirements of the Silver-Blaze property~\cite{PhysRevLett.86.592,PhysRevLett.91.222001}, where all the quantities (except charged meson masses) are independent of $\mu_I$, for $\mu_I<\mu_I^c$ at $T=0$~\cite{Adhikari:2018cea}.

However, for the finite temperature case, the Matsubara frequency summation introduces Fermi-Dirac distribution terms into the one-loop effective potential, 
so the critical value is no longer determined by $\mu_I^c = \mathring{M}_\pi/2$ in Eq.~(\ref{GapEqDelta0}).

\begin{figure}[htbp]
\includegraphics[width=0.88\linewidth]{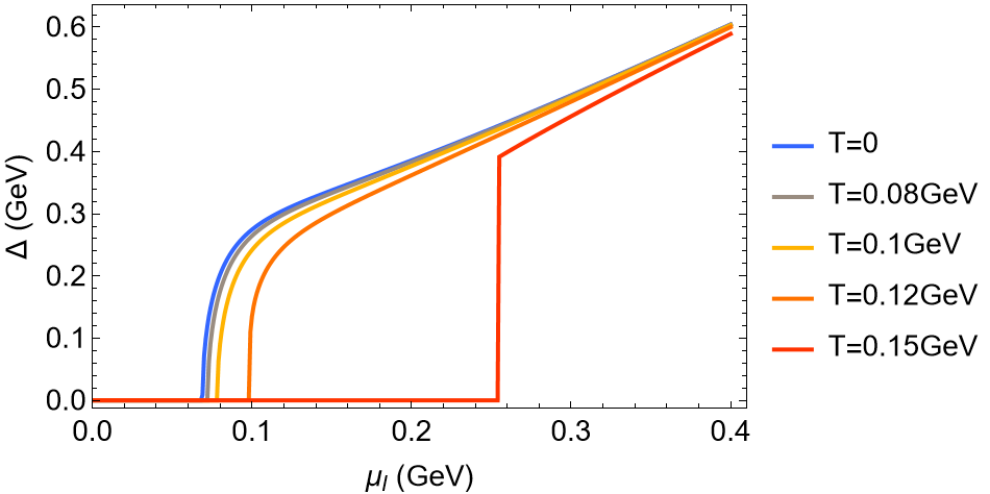}
\caption{The pion condensate $\D$ as a function of isospin chemical potential at different temperatures.}
\label{fig:pion condensation with mu}
\end{figure}

We exhibit the pion condensate $\Delta$ as a function of isospin chemical potential at different temperatures in Fig.~\ref{fig:pion condensation with mu}. At low temperatures, the gap $\D$ is not sizable until $\mu_I$ hits the critical value, and above this point it starts to increase without any jumps. Thus, the phase transition to the pion superfluid phase is of second-order. At $T=0$, the critical isospin chemical potential is exactly given by $\mu^c_I=\mathring{M}_\pi/2$ as analytically derived above. Temperature effects shift the phase-transition point to the right due to the inhibition from quark thermal excitations.

The nature of the phase transition shows a qualitative change as the temperature increases. As shown in Fig.~\ref{fig:pion condensation with mu}, at $T=0.15\, \text{GeV}$, the order parameter no longer varies continuously with $\mu_I$, but instead exhibits a sudden jump from zero to a finite value at $\mu_I^c\approx0.255,\text{GeV}$. Such a discontinuity signals the emergence of first-order behavior in the pion-superfluid transition within our mean-field treatment. Similar behavior has been reported in previous studies of QCD$_I$~\cite{Ferreira:2025zeu,Kamikado:2012bt,Zhang:2006gu,Kouno:2011zu,Sasaki:2010jz,Kovensky:2024oqx,Adhikari:2018cea} and two-color QCD~\cite{Kogut:2001if,Chandrasekharan:2010ik,Kogut:2002kj,Splittorff:2002xn,Strodthoff:2011tz,Dunne:2003ji,Wirstam:2002be}. Though this first-order phase transition is possiblly a mean-field artifact, in order to further understand its origin, we give a Ginzburg--Landau analysis in Appendix~\ref{ap:Laudau theory}.

In dense regime where the pion superfluidity is developed, Fig.~\ref{fig:pion condensation with mu} shows that the pion condensates with any temperatures exhibit approximately linear dependencies on $\mu_I$: $\D\sim {\cal C}\m_I$, regardless of temperatures. The coefficient ${\cal C}$ is given by ${\cal C} =  \sqrt{24g^2/\lambda_2}$ which is indeed independent of temperatures, as derived in Eq.~(\ref{DeltaMuIDep}).

\subsection{Phase diagram}
\label{sec:PhaseDiagram}

\begin{figure}[htbp]
\centering
\vspace{0.05cm}
\includegraphics[width=0.9\linewidth]{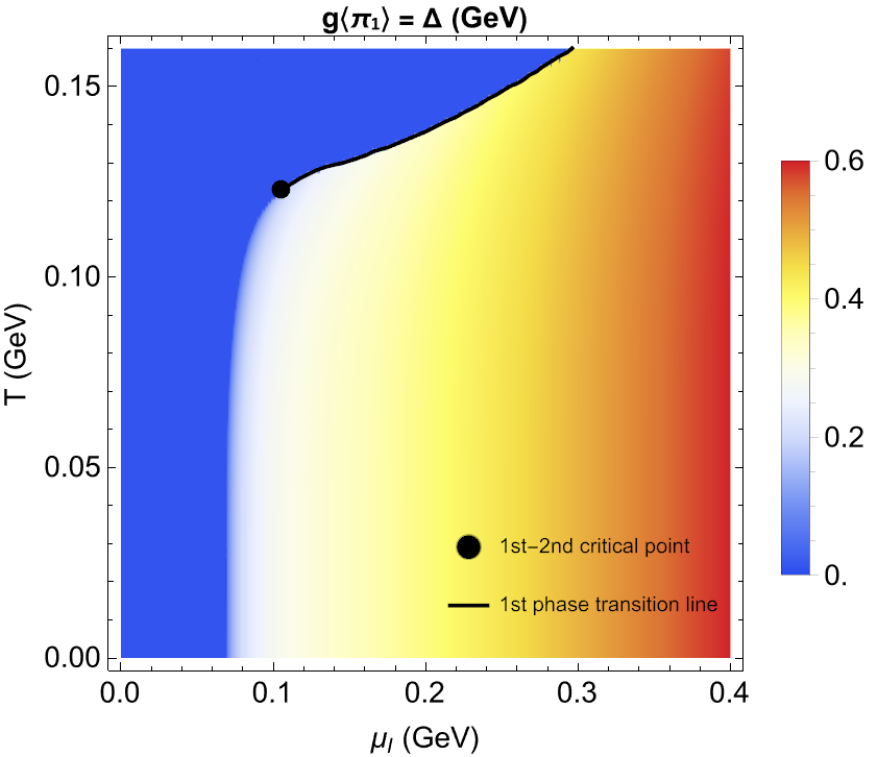}

\vspace{0.1cm}

\vspace{0.05cm}
\includegraphics[width=0.9\linewidth]{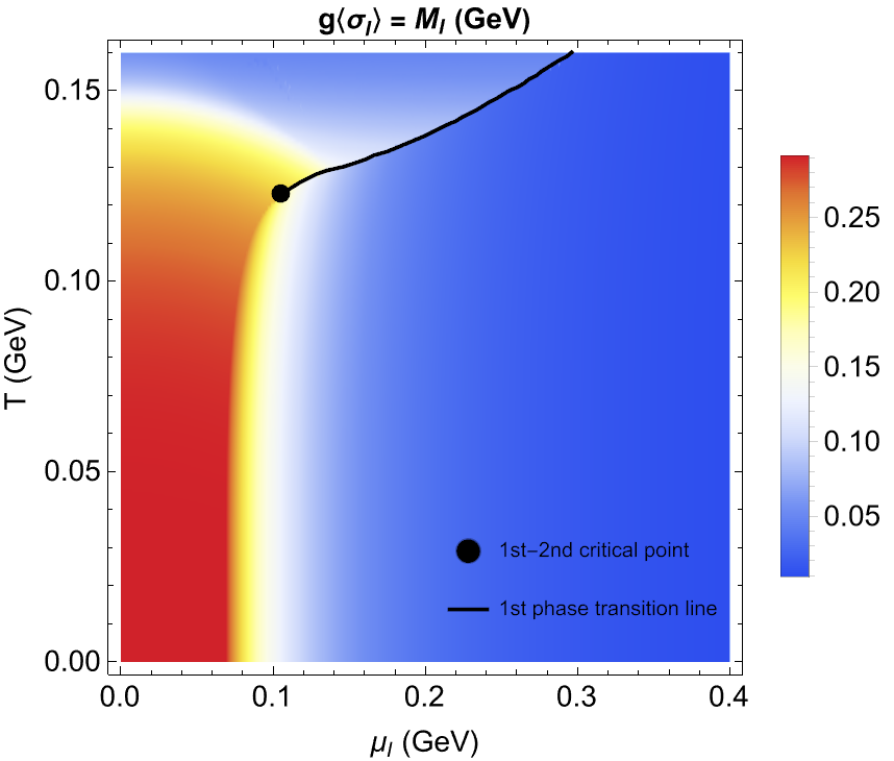}

\vspace{0.1cm}

\vspace{0.05cm}
\includegraphics[width=0.9\linewidth]{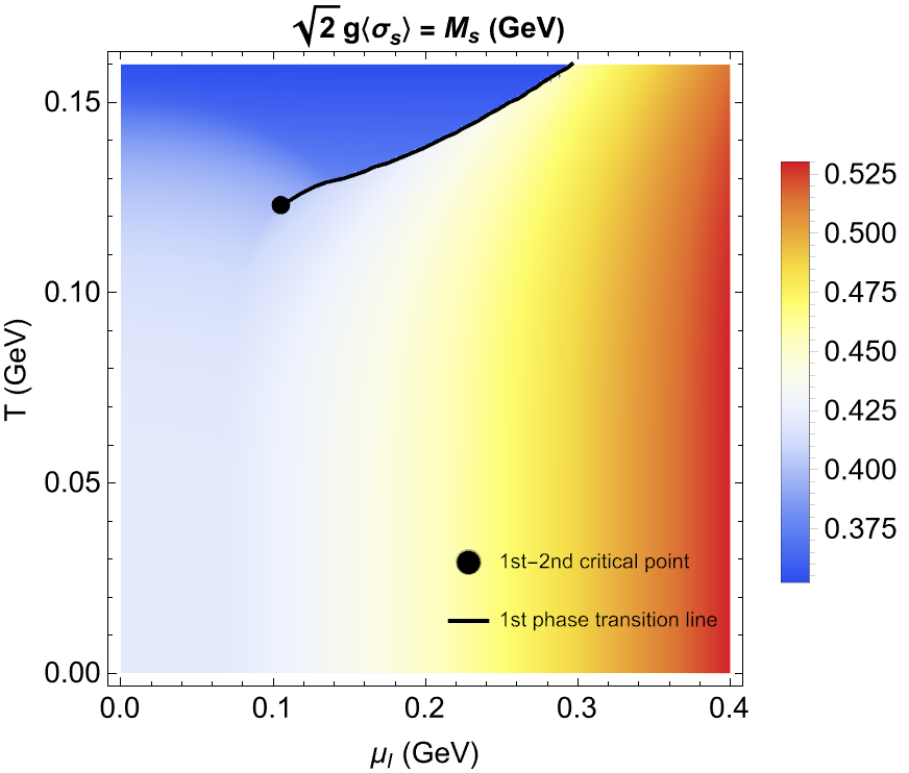}
\caption{Order parameters $\Delta$ (top), $M_l$ (middle), and $M_s$ (bottom) as functions of temperature and isospin chemical potential. The first-order phase boundary is indicated by black curves.}
\label{fig:full phase}
\end{figure}

Figure~\ref{fig:full phase} illustrates the two-dimensional contour plots of the constituent light quark mass $M_l$, strange quark mass $M_s$, and pion condensate $\Delta$ in the $T$-$\mu_I$ plane. At low temperature and low isospin chemical potential, the system stays in a normal phase where the pion condensate vanishes, and both light and strange quarks maintain large dynamical masses due to spontaneous chiral symmetry breaking. As isospin chemical potential increases beyond the critical value $\mu_I^c$ at low temperatures, $\Delta$ acquires a non-zero expectation value, marking the system entering pion superfluid phase. Concurrently, the light quark mass experiences a substantial reduction as $M_i \sim ( \mathring{M}_l \mathring{M}_\pi^2/4)\mu_I^{-2}$ at large $\mu_I$, as derived in Eq.~(\ref{MlMuIDep}). Meanwhile, the strange quark mass exhibits an increment owning to the nonzero anomaly effects ($K>0$). The critical point where the second-order transition changes to the first one is located at $(T,\mu_I) \approx (0.122\, \text{GeV}, 0.103\, \text{GeV})$. All the three order parameters show discontinuous jumps along this black curve.

The discontinuous drop of $M_l$ can be attributed to the discontinuous change of $\Delta$, since these quantities involve only the light-flavor sector. Meanwhile, the jump of $M_s$ along the black curve is associated with the nonzero $K$ term, i.e., the $U(1)_A$ anomaly effects that couple the three flavors. A detailed discussion is provided in Appendix~\ref{ap:GapEquation}.

\section{meson spectra}
\label{sec:meson}

In this section, we present one of our main results on the meson mass spectrum in QCD$_I$ medium, which provides a useful probe of the underlying phase structure from the perspective of chiral symmetry. Previous studies have investigated the meson spectrum in the pion-condensed phase within the three-flavor NJL model~\cite{Xia:2013caa}. 
in which mesonic excitations are described as quark-antiquark bound states. Meanwhile, it should be emphasized that our present quark-meson model consists of explicit meson degrees of freedom as well as quark ones, which is capable of providing a comprehensive description of the mass spectrum from low density to high density.

\subsection{Notes}
\label{sec:Note}

As illustrated in Appendix~\ref{ap:meson mass}, the meson mass is defined by an energy at which the corresponding propagator matrix has its pole at the rest frame. The meson propagator-inverse matrix generally takes the form of
\begin{equation}
D^{-1}(p)_{\alpha\beta}= D_0^{-1}(p)_{\alpha\beta}-m^2_{\text{tree},\alpha\beta}-\Pi(p)_{\alpha\beta}\ ,
\end{equation}
where the subscripts $\alpha,\beta$ represent the meson fields. $D_0^{-1}(p)$ is the ``zero-th order'' propagator-inverse matrix which may include the chemical-potential effects. And the second term $m_{\rm tree}^2$ corresponds to the tree-level-mass matrix. The last term, $\Pi(p)$, is the one-loop self-energy matrices which in general can be expresses as
\begin{equation}
\Pi(p)_{\alpha\beta} = -g^2\int_k
\mathrm{Tr}
\left[
(\Gamma\lambda)_\alpha^\dagger S_q(k+p)(\Gamma\lambda)_{\beta} S_q(k) \right]\ ,
\end{equation}
where the symbol ``Tr'' stands for the trace operator for the Dirac, color and flavor spaces, and $\int_k$ represents the Matsubara summations as well as the three-dimensional momentum integrals. The matrix $(\Gamma\lambda)_{\alpha}$ represents a product of the appropriate Dirac and Gell-Mann matrices. The explicit expressions for $m^2_{\text{tree}}$ and $\Pi(p)$ are presented in Appendix~\ref{ap:meson mass}. We note that the propagator matrix can be divided into several blocks, as listed in Table~\ref{tab:meson}.

In this section, we adopt the following particle basis:
\begin{eqnarray}
&& \pi_0=\pi_3\ , \ \ \pi_\pm = \frac{\pi_1\mp \mi\pi_2}{\sqrt{2}}\ , \ \ \eta_l=\pi_l\ , \ \ \eta_s = \pi_s\ ,\nonumber\\
&& a_{0,0}=\sigma_3\ , \ \  a_{0,\pm} = \frac{\sigma_1\mp \mi\sigma_2}{\sqrt{2}}\ , \nonumber\\
&& K_\pm = \frac{\pi_4  \mp\mi \pi_5}{\sqrt{2}}\ , \ \ \kappa_\pm = \frac{\sigma_4 \mp\mi \sigma_5}{\sqrt{2}}\ , \label{MesonName}
\end{eqnarray}
with which $\sigma_l$ and $\sigma_s$ are employed as they are. Although the anomalous term induces mixing between $\bar{l}l$ and $\bar{s}s$ components ($l=u,d$), we refer to $\sigma_l$ ($\eta_l$) and $\sigma_s$ ($\eta_s$) as the lower and higher mass eigenstates. We adopt these manners and Eq.~(\ref{MesonName}) even in the superfluid phase despite further complicated mixings.

\subsection{Mass spectra at $T=0$}
\label{sec:ZeroTMass}

Here, first we show the meson mass spectra at $T=0$ as the simplest examination.

\begin{figure}[htbp]
\centering
\vspace{0.05cm}
\includegraphics[width=0.9\linewidth]{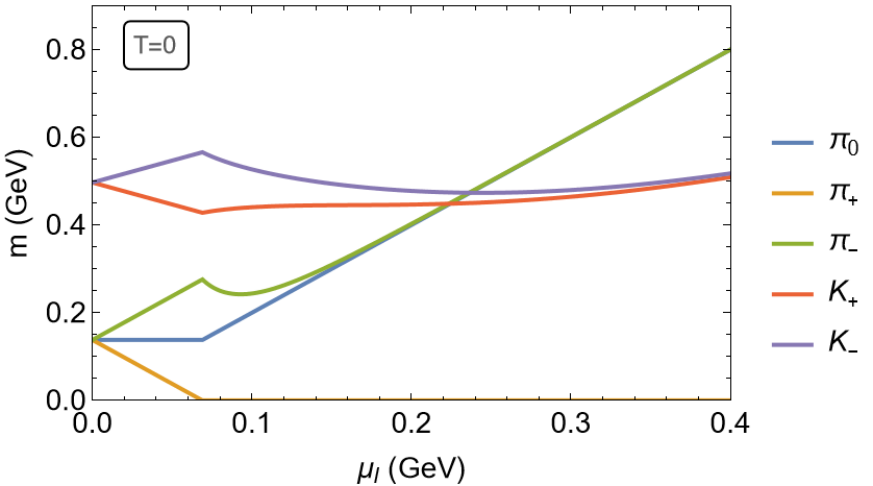}

\vspace{0.1cm}

\vspace{0.05cm}
\includegraphics[width=0.9\linewidth]{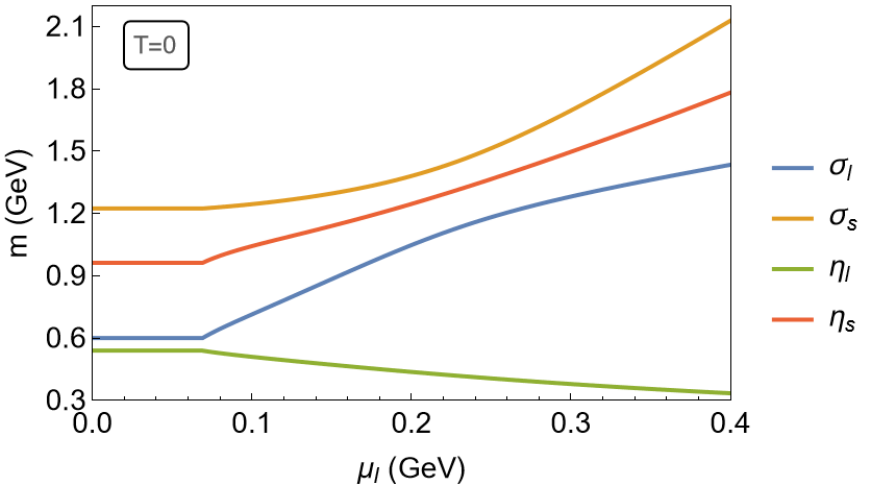}

\vspace{0.1cm}

\vspace{0.05cm}
\includegraphics[width=0.9\linewidth]{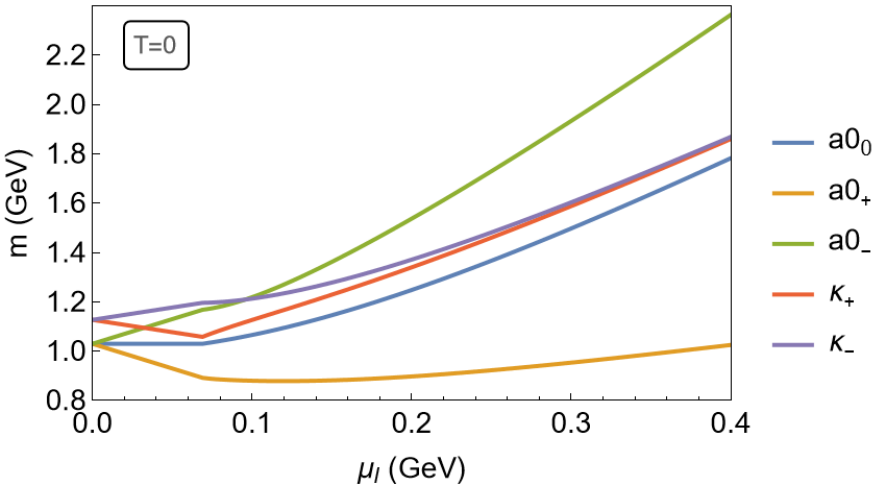}
\caption{Meson mass spectra as a function of isospin chemical potential at $T=0$.}
\label{fig:meson mass T=0}
\end{figure}

Depicted in Fig~~\ref{fig:meson mass T=0} are the resultant mass spectra as a function of the isospin chemical potential. In the normal phase but with finite $\mu_I$ ($\mu_I \leq \mu_I^c$ with $\mu_I^c=\mathring{M}_\pi/2$), the explicit breaking of $SU(2)$ isospin symmetry induces a universal mass splitting for all mesons carrying a non-zero isospin charge. Consequently, the charged states ($\p_{\pm},a_{0\pm},K_{\pm},\k_{\pm}$) exhibit linear $\mu_I$ dependencies accordingly to their isospin charges as explained above Eq.~(\ref{MeanField}), while the isospin-neutral states remain unaffected. 

In the pion superfluid phase ($\mu_I^c < \mu_I$), the $\pi_+$ mass vanishes exactly, acting as the Goldstone boson associated with the spontaneous breaking of the residual $U(1)_{\l_3}$ symmetry. Concurrently, the chiral condensate $\langle\sigma_l\rangle$ partially turns into the pion condensate, which generates a off-diagonal background in $\langle M\rangle$ that profoundly modifies the remaining mass spectrum. In particular, the pronounced mixing among scalar and pseudoscalar mesons yields complex nonlinear mass trajectories, reflecting the intricate relation between the superfluid medium and the partial restoration of chiral symmetry. 

One of the noteworthy characteristics in the pion superfluid phase is the strictly linear mass growth of the massive excitation ($\p_0$ in our model) with increasing $\mu_I$: $M_{\pi_0}=2\mu_I$. Such an exact mass formula for the massive NG boson was derived in Refs.~\cite{PhysRevLett.110.011602,PhysRevLett.111.021601} within a general consideration. Although our present analysis is based on the $N_f=2+1$ QCD$_I$, which would possess a rather complicated symmetry structure, one can expect that the arguments in these references can apply to our consideration.

\begin{figure}[htbp]
\centering
\vspace{0.05cm}
\includegraphics[width=0.9\linewidth]{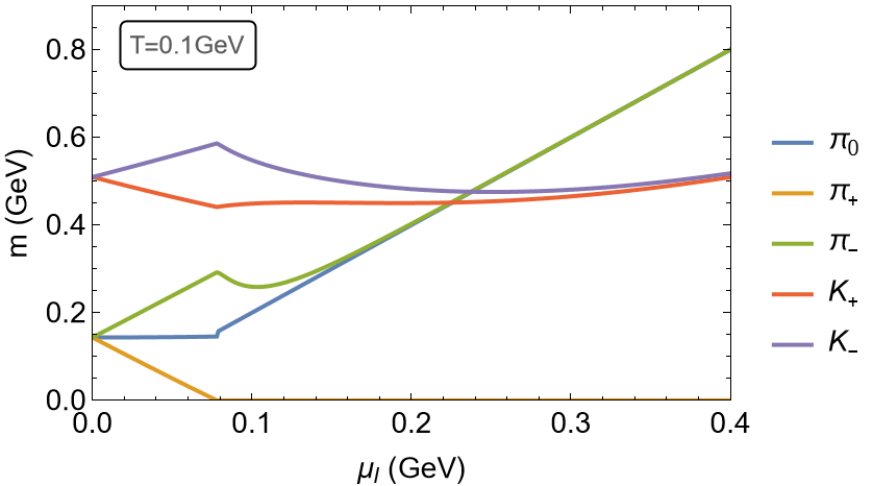}

\vspace{0.1cm}

\vspace{0.05cm}
\includegraphics[width=0.9\linewidth]{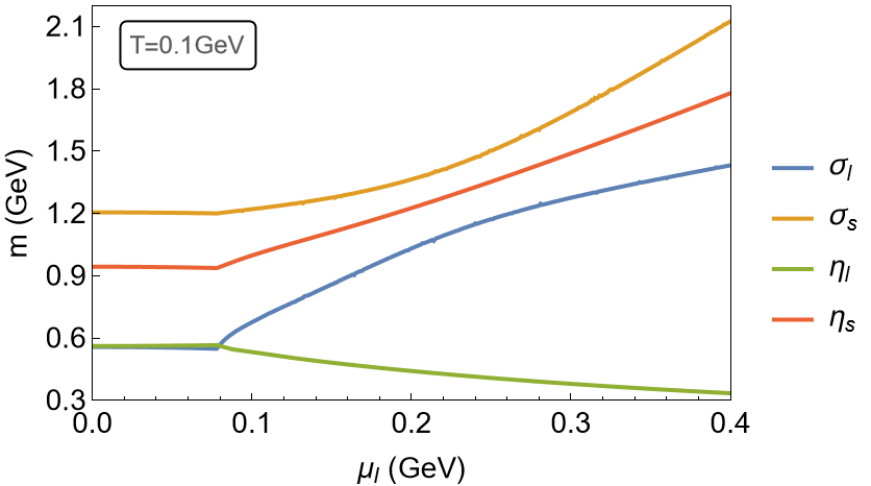}

\vspace{0.1cm}

\vspace{0.05cm}
\includegraphics[width=0.9\linewidth]{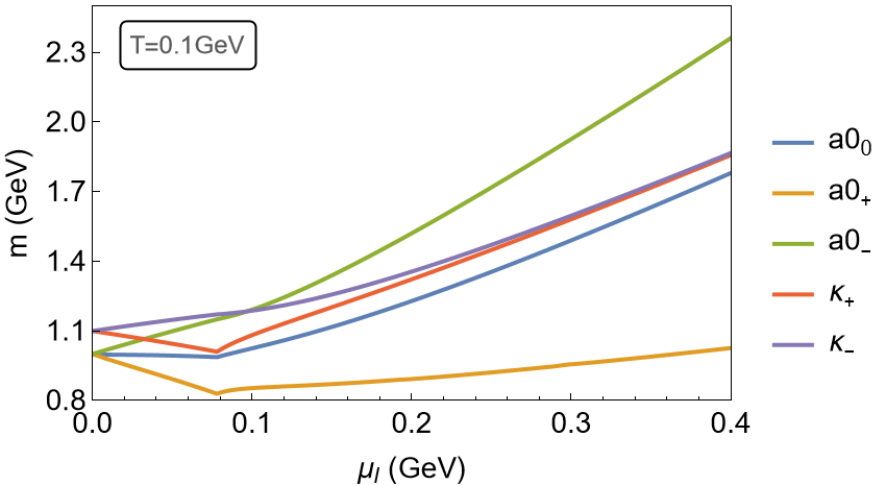}
\caption{Meson mass spectra as a function of isospin chemical potential at $T=0.1\, \text{GeV}$.}
\label{fig:meson mass T=0.1}
\end{figure}

\begin{figure}[htbp]
\centering
\vspace{0.05cm}
\includegraphics[width=0.85\linewidth]{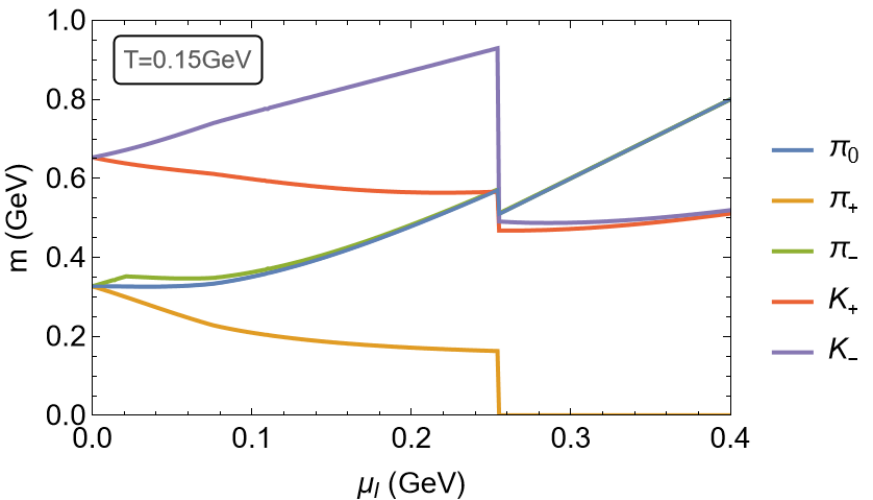}

\vspace{0.1cm}

\vspace{0.05cm}
\includegraphics[width=0.85\linewidth]{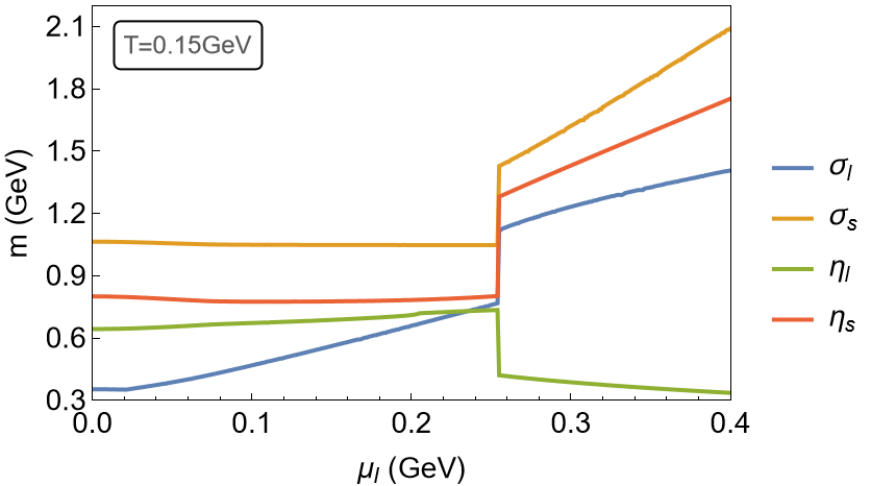}

\vspace{0.1cm}

\vspace{0.05cm}
\includegraphics[width=0.85\linewidth]{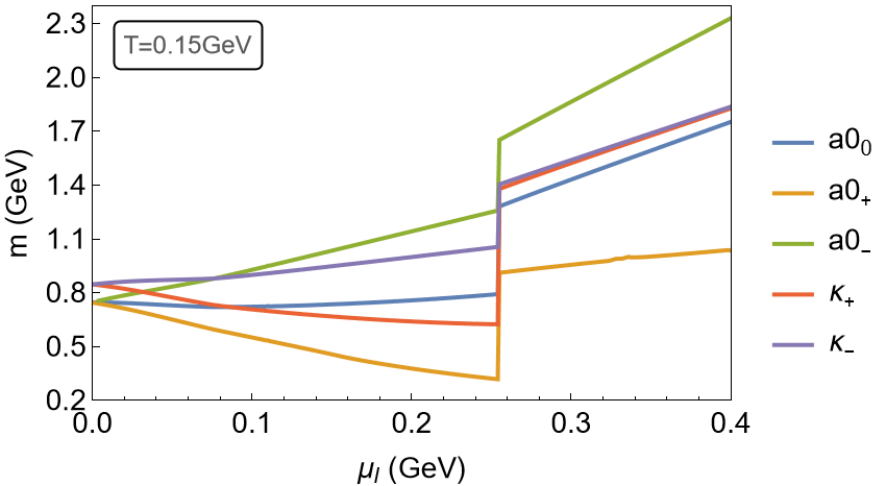}
\caption{Meson mass spectra as a function of isospin chemical potential at $T=0.15\, \text{GeV}$. The jumps are }
\label{fig:meson mass T=0.15}
\end{figure}

Another remarkable finding in the superfluid phase is the mass reduction of $\eta_l$, owing to the mixings from $\eta_s$, $a_0^+$ and $a_0^-$, whereas all the other meson masses (except the NG mode $\pi_+$) increase with $\mu_I$. Such a peculiar behavior claims that the second-lowest mode in the superfluid phase is $\eta_l$; at certain $\mu_I$, $\eta_l$ mass gets lighter than $\pi_0$ and $\pi_-$ masses. This mass inversion is derived from the linear representation of the quark-meson model incorporating parity partners which cannot be describe by the ChPT. It should be noted that similar mass hierarchy was discussed in the baryon superfluid phase in two-color QCD, with which the lattice simulation indeed indicated it consistently~\cite{Suenaga:2022uqn,Suenaga:2025sln,Iida:2026dmw}.

Although the non-strange and hidden-strange meson masses exhibit various significant behaviors, the open-strange meson, $K_\pm$ and $\kappa_\pm$, masses are rather simple. In the vicinity of $\mu_I\sim \mu_I^c$, the respective charged partners once split due to the $SU(2)$ isospin breaking. In the large $\mu_I$ regime, the masses again tend to degenerate reflecting the partial chiral restoration, as discussed in Appendix~\ref{ap:TreeLevelMass}. In this limit, the anomaly effects on those open-strange mesons are suppressed owing to the chiral restoration.

\subsection{Mass spectra at finite temperature}
\label{sec:FiniteTMass}

Basic properties of the $N_f=2+1$ meson mass spectra in the pion superfluid phase in QCD$_I$ have been delineated in Sec.~\ref{sec:ZeroTMass}. In this subsection, we extend the analysis to finite temperature systems to gain insights into the thermal effects on the masses.

Figure~\ref{fig:meson mass T=0.1} indicates the resultant mass spectra at $T=0.1$ GeV. Compared to the result at $T=0$ in Fig.~\ref{fig:meson mass T=0}, the stable behaviors in the normal phase are slightly extended to larger $\mu_I$, reflecting delay of the phase transition due to the inhibition from thermal quarks, consistently with $\Delta$ in Fig.~\ref{fig:pion condensation with mu}. Despite such a tiny deviation, qualitatively the overall tendency of the mass evolution with $\mu_I$ remains consistent with the zero-temperature case. That is, fundamental responses of the meson masses to the isospin density is robust against moderate thermal fluctuations.

At finite temperature, thermal fluctuations prevent from forming  chiral condensate, which leads to the partial restoration of chiral symmetry. Consequently, the mass splitting between chiral partners, which is maximized in the vacuum, begins to decrease. This mechanism is explicitly reflected in the numerical results; compared to the $T=0$ case, the overall masses of the pseudoscalar mesons exhibit an upward shift, whereas those of the scalar mesons decrease at $\mu_I=0$.

At sufficiently high $T$, the pion-superfluid transition exhibits first-order behavior within the present approximation, as indicated by the discontinuous change of the pion condensate shown in Fig.~\ref{fig:pion condensation with mu}. As a consequence, the meson masses also show discontinuities across the transition, as illustrated for $T=0.15$ GeV in Fig.~\ref{fig:meson mass T=0.15}. We, however, emphasize that these discontinuities and the associated  first-order transition itself may be artifacts of the mean-field approximation. Resolving their physical status requires analyses beyond mean field. 

Because $T=0.15$ GeV is close to the pseudo-critical temperature for chiral restoration, the dynamical light-quark mass $M_l$ is already reduced sizably at $\mu_I=0$. Hence, particularly the light-flavor-meson masses exhibit significant sensitivity against $\mu_I$, even for the isospin-neutral mesons. Besides, due to the approximate mass degeneracies between the chiral partners, for instance, one can see $M_{\sigma_l}\sim M_\pi\sim 0.33$ GeV at $\mu_I=0$. This large $M_{\pi}$ together with the large sensitivity against $\mu_I$ results in a considerably large mass of $\pi_+$: $M_{\pi_+}\sim 0.15$ GeV just below the critical point $\mu_I^c \sim0.255$ GeV. The masses of strange mesons in the normal phase remain comparably stable compared to purely light meson masses. This disparity naturally arises from the larger dynamical strange-quark mass. Then, the $\pi_+$ mass exhibits a discontinuous change across the transition and reaches zero in the pion-condensed phase, as shown in the top panel of Fig.~\ref{fig:meson mass T=0.15}. 

Before closing this section, we comment on the $\pi_0$ mass in the pion superfluid phase. Remarkably, the relation $M_{\pi_0}=2\mu_I$ remains valid even at finite temperature, despite the substantial modification from thermal fluctuations. We provide an analytical proof of this relation in Appendix~\ref{ap:PionMass}. This observation suggests that the general argument on the massive NG boson in Refs.~\cite{PhysRevLett.110.011602,PhysRevLett.111.021601} can be applied to finite temperature system~\cite{PhysRevD.91.056006}.

\section{Sound velocity}
\label{sec:EoS}

The pion-condensed phase is not only characterized by its order parameter, but also has influence on the thermodynamic properties of the isospin medium. In particular, the isospin density and its susceptibility characterize the response of the system to the isospin chemical potential~\cite{PhysRevLett.86.592,Lu:2019diy,He:2005nk}, while the sound velocity provides a direct measure of the stiffness of the equation of state~\cite{Chiba:2024cny,Suenaga:2025sln}.

The pressure is readily obtained by $P=-\O$ from Eq.~(\ref{eq:thermal potential}), which allows us to evaluate the isospin charge density $n_I=\partial P/\partial\mu_I$ and entropy density $s=\partial P/\partial T$ as
\begin{eqnarray}
n_I &=& \frac{4\D^2}{g^2}\m_I+4N_c\sum_{\xi={\rm p,a}} \int_{\vec{p}}\frac{\eta_\xi(E_l-\eta_\xi \m_I)}{\epsilon_\xi}f(\epsilon_\xi ) \ , \nonumber\\
s &=&4N_c \int_{\vec{p}} \Bigg\{ \sum_{\xi={\rm p,a}} \left[\frac{\epsilon_{\xi}}{T}f(\epsilon_\xi )  +\mln\left(1+\me^{-\frac{\epsilon_\xi}{T}} \right) \right] \nonumber\\
&&  +\frac{E_s}{T} f(E_s) +\mln(1+\me^{-\frac{E_s}{T}}) \Bigg\} \ . \end{eqnarray}
The energy density $\epsilon$ is readily evaluated by $\epsilon = -P+T s +\mu_I n_I$.

\begin{figure}[htbp]
\centering
\vspace{0.05cm}
\includegraphics[width=0.9\linewidth]{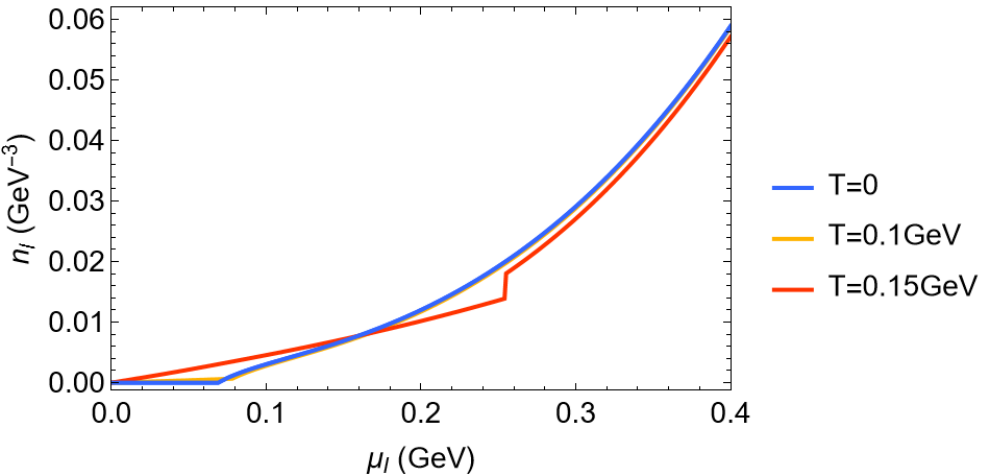}

\vspace{0.1cm}

\vspace{0.05cm}
\includegraphics[width=0.9\linewidth]{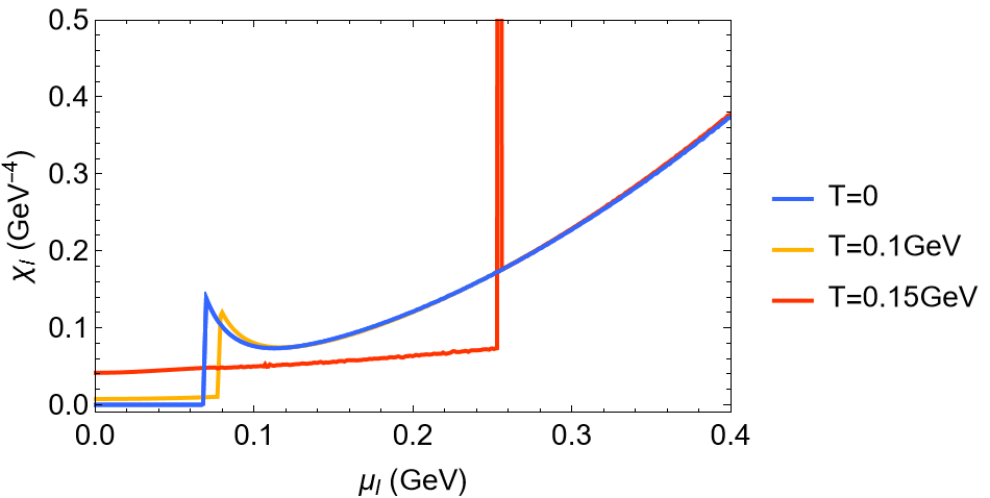}

\vspace{0.1cm}

\includegraphics[width=0.9\linewidth]{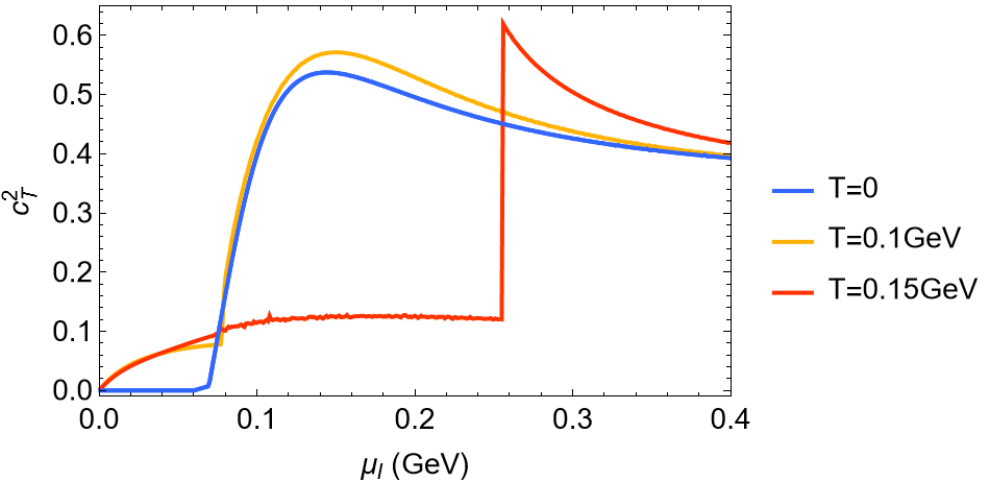}
\caption{$\mu_I$ dependencies of the isospin density $n_I$ (top), the isospin susceptibility $\chi_I=\partial n_I/\partial \mu_I$ (middle), and the squared isothermal sound velocity $c_T^2$ (bottom).}
\label{fig: state equation}
\end{figure}

In Fig.~\ref{fig: state equation}, we depict the evolution of the isospin density $n_I$ (top), the isospin susceptibility $\c_I=\partial n_I/\partial \mu_I$ (middle panel). Since on the lattice the sound velocity is typically measured along isothermal trajectories, we also exhibit the isothermal sound velocity $c_T^2 = (\partial P / \partial \epsilon)_{T}$ in the bottom panel in Fig.~\ref{fig: state equation}. At zero temperature, the system strictly adheres to the Silver-Blaze property~\cite{PhysRevLett.86.592,PhysRevLett.91.222001}; for $\mu_I$ below the critical point $\mu_I^c$, all the three quantities remain exactly zero. At the critical point the continuous increase in $n_I$ signals the appearance of pion condensate, and $n_I$ monotonically grows in the superfluid phase. Concurrently, the susceptibility $\chi_I$ exhibits a jump reflecting the nonanalytic behavior of the second derivative of the thermodynamic potential at $\mu_I=\mu_I^c$, that is the characteristic of a second-order phase transition. The isothermal squared sound velocity $c_T^2$ continuously evolves in the superfluid phase making a peak structure around $\mu_I\sim  2\mu_I^c$.

At finite temperature, the thermal excitations  generate a finite isospin density even below the critical $\mu_I$. Macroscopically, this is manifested as a gradual and continuous rise in $n_I$, Accordingly, $\chi_I$ and $c_T^2$ are also finite even in the normal phase. Despite these distinctions, their qualitative behaviors are similar to those at zero temperature, as long as $T$ is lower. When temperature is further increased, our model with the mean-filed approximation yields the first-order transition. Consequently, $n_I$ exhibits a clear discontinuity at the critical point, which translates into a Dirac $\delta$-function-like singularity in the isospin susceptibility. 

The isothermal sound velocity $c_T^2$ in the normal phase exhibits a broad bump at $T=0.15$ GeV. Around the critical point, it shows a discontinuous enhancement and then decreases monotonically in the superfluid phase. This contrasting behaviors are induced by the first-order transition. Hence, although the first-order transition could be an artifact of the present mean-field approach, it is demonstrated that the isothermal sound velocity provides complementary information on the phase structure.

Here, we comment on the reduction of $\chi_I$ just above the critical point at lower temperatures. This reduction reflects the weak response against density fluctuations after the onset of pion superfluidity. Such a stable property is also reflected in a suppression of the corresponding compressibility $\kappa_T = \chi_I/n_I^2$. Those behaviors indicate a transient stiffening of the equation of state, providing the physical origin of the sound-velocity peak that emerges in the superfluid phase as will be explained below.

\begin{figure}[htbp]
\includegraphics[width=1\linewidth]{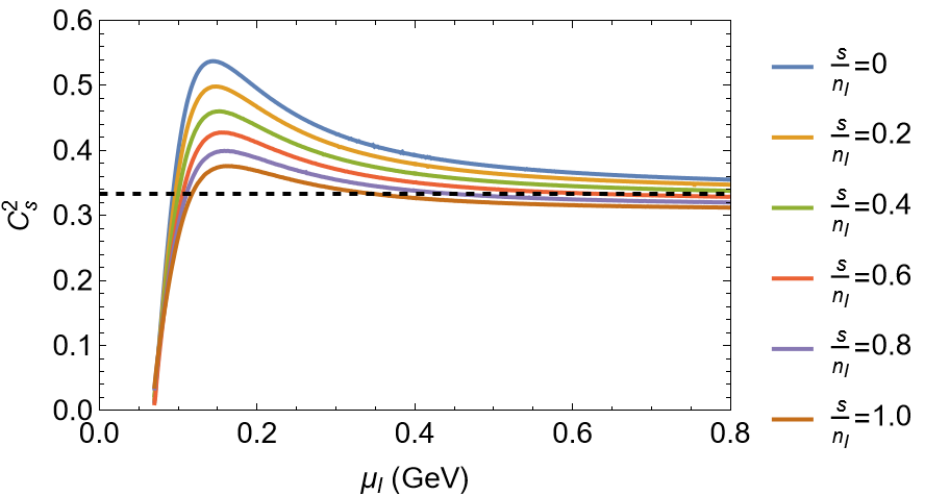}
\caption{Squared sound velocity $c_s^2$ as a function of isospin chemical potential with several $s/n_I$.}
\label{fig:cs}
\end{figure}

For completeness, we also depict $\mu_I$ dependencies of the squared sound velocity $c_s^2 = (\partial P / \partial \epsilon)_{s/n_I}$ along several iso-entropic trajectories, in Fig.~\ref{fig:cs}.  At intermediate densities, following the onset of pion condensate, $c_s^2$ exhibits a rapid increase. This behavior indicates a substantial stiffening of the EoS~\cite{Baym:2017whm}, driven by the strongly repulsive nature of the effective interactions within the dense pion superfluid. The $c_s^2$ for small $s/n_I$ exceeds the conformal limit $\bar{c}_s^2=1/3$ to exhibit a pronounced peak slightly above $\mu_c \sim \mathring{M}_\pi$. As $s/n_I$ increases, thermal fluctuations gradually soften the EoS resulting in the suppression of the peak height.

In extreme high-density regime, the thermodynamics will be dominated by the kinetic energy of effectively massless particles in such a way as to restore scale invariance. Indeed, our extrapolated results demonstrate that all $c_s^2$ curves converge on the conformal limit $\bar{c}_s^2 = 1/3$, which illustrates the transition from a strongly interacting matter to a weakly interacting ultra-relativistic conformal gas.

\section{role of axial anomaly}
\label{sec:anomaly}

By means of the FRG method, it is discussed that the $U(1)_A$ anomaly effects on hadrons in medium can be enhanced, before being suppressed due to the Debye screening of electric gluons in extremely high temperature and density~\cite{Fejos:2025oxi,Fejos:2026tyd}. In this regard, examinations of responses against changing the anomaly effects are expected to provide us with useful information on such modifications. For this reason, in this section we further study the dependence on the KMT-interaction strength $K$.

\begin{table}[h]
\centering
\begin{ruledtabular}
\begin{tabular}{c||cc|cc}
KMT coupling & $m_M^2$ [GeV$^2$]  & $\l_2$ & $\mathring{M}_{\w_l}$ [GeV] & $\mathring{M}_{\w_s}$ [GeV] \\
\hline
$0.25\bar{K}$ & 0.2162 & 221.6 & 0.397 & 0.681 \\
$0.5\bar{K}$ & 0.0979 & 194.7 & 0.488 & 0.763 \\
$0.75\bar{K}$ & -0.0204 & 167.8 & 0.524 & 0.863 \\
$\bar{K}$ & -0.1386 & 140.9 & 0.540 & 0.962 \\
$1.25\bar{K}$ & -0.2570 & 87.1 & 0.548 & 1.143 \\
\end{tabular}
\end{ruledtabular}
\caption{Parameters under different KMT coupling. The other parameters $g=3.15$, $h_l=1.759\times10^{-3}\, \text{GeV}^3$,$h_s=5.407\times10^{-2}\, \text{GeV}^3$ are unchanged, while we fix $\lambda_1=0$ in this section for simplicity. }
\label{tab:kmt coeffiecnt}
\end{table}

\begin{figure}[htbp]
\includegraphics[width=1\linewidth]{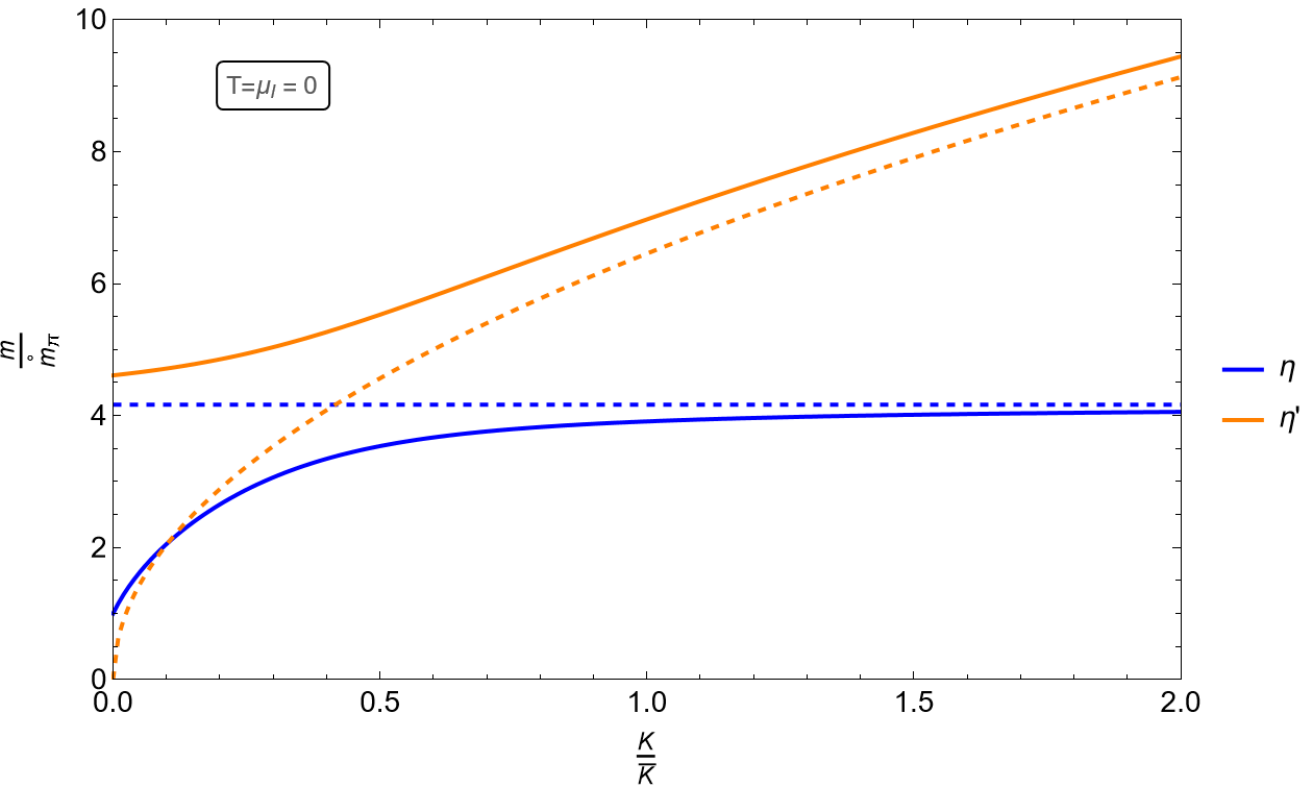}
\caption{The vacuum $\w$ and $\eta'$ masses, $\mathring{M}_{\eta_l}$ and $\mathring{M}_{\eta_s}$, as a function of normalized KMT coupling at $T=\m_I=0$, scaled by vacuum pion mass $\mathring{M}_\pi$. }
\label{fig:vacuum eta mass}
\end{figure}

\subsection{$K$ dependence of $\eta$ masses}
\label{sec:EtaMass}

Before showing the results, here we argue the $K$ dependencies of $\eta$ and $\eta'$ masses, i.e., $\mathring{M}_{\eta_l}$ and $\mathring{M}_{\eta_s}$.\footnote{Recall that $\eta$ and $\eta'$ masses are expressed by $\mathring{M}_{\eta_l}$ and $\mathring{M}_{\eta_s}$ in our notation.}

We define $\bar{K}=1.20$ GeV to refer to the real-world value of KMT coupling tabulated in Table~\ref{tab:Parameter}. Then we consider several values of the anomaly strength. For each choice of $K$, the remaining model parameters are refitted for which the vacuum pion mass, kaon mass, the corresponding decay constants, and the chiral condensates remain unchanged. The refitted parameters can be found in Table~\ref{tab:kmt coeffiecnt}. 

Such a refitting procedure is based on an effective-model spirit, where variations of the axial-anomaly interaction are expected to be accompanied by corresponding changes in the other low-energy effective parameters, while maintaining the same vacuum phenomenology. As naturally expected, with the different KMT coupling, quantities strongly related to the axial anomaly such as the two $\w$ meson masses deviate from their physical values. Figure~\ref{fig:vacuum eta mass} illustrates the vacuum masses of the $\eta$ and $\eta'$ ($\mathring{M}_{\eta_l}$ and $\mathring{M}_{\eta_s}$) as a function of the normalized $K/\bar{K}$. The artificial $U_A(1)$ symmetry ($K=0$) enforces a strict mass degeneracy between the $\eta_l$ and pion. The finite KMT interaction lifts this degeneracy and induces a highly asymmetric evolution; the heavy $s\bar{s}$-dominant $\eta'$ experiences a drastic mass increase, whereas the light $l\bar{l}$-dominant $\eta$ remains stable for $K/\bar{K} \gtrsim 1$. 

More quantitatively, in terms of the inputs $\mathring{M}_\pi$, $\mathring{M}_K$, $f_\pi$ and $f_K$, the mass-matrix elements for $\eta_l$ and $\eta_s$ system is expressed as
\begin{eqnarray}
m_{\pi_l\pi_l}^2 &=& \mathring{M}_\pi^2 + 4(2f_K-f_\pi)K \ , \nonumber\\
m_{\pi_s\pi_s}^2 &=&  \frac{2f_K \mathring{M}_K^2-f_\pi \mathring{M}_\pi^2 + 2f_\pi^2 K}{2f_K-f_\pi} \ , \nonumber\\
m_{\pi_l\pi_s}^2 &=& 2\sqrt{2}f_\pi K \ ,
\end{eqnarray}
from Eqs.~(\ref{MEtaL}) -~(\ref{MEtaS}). After diagonalizing, in a large $K$ limit the resultant mass eigenvalues read
\begin{eqnarray} 
\mathring{M}_{\eta_l} &\sim& \left(\frac{4(2f_K^2-f_\pi f_K)\mathring{M}_K^2 + (3f_\pi^2-4f_\pi f_K)\mathring{M}_\pi^2}{8f_K^2-8f_K f_\pi + 3f_\pi^2}\right)^{1/2}\ , \nonumber\\
\mathring{M}_{\eta_s} &\sim& \left(\frac{2(8f_K^2-8f_K f_\pi + 3f_\pi^2)}{2f_K-f_\pi}K \right)^{1/2}\ ,
\end{eqnarray}
which are denoted by dashed lines in Fig.~\ref{fig:vacuum eta mass}. These mass formulas explicitly illustrate that the squared $\eta'$ mass depends linearly on $K$, whereas the $\eta$ mass remains nearly stable~\cite{Hatsuda:1994pi,tHooft1976}.

\subsection{Anomaly effects in medium}
\label{sec:AnomalyT}

\begin{figure}[htbp]
\centering
\includegraphics[width=0.9\linewidth]{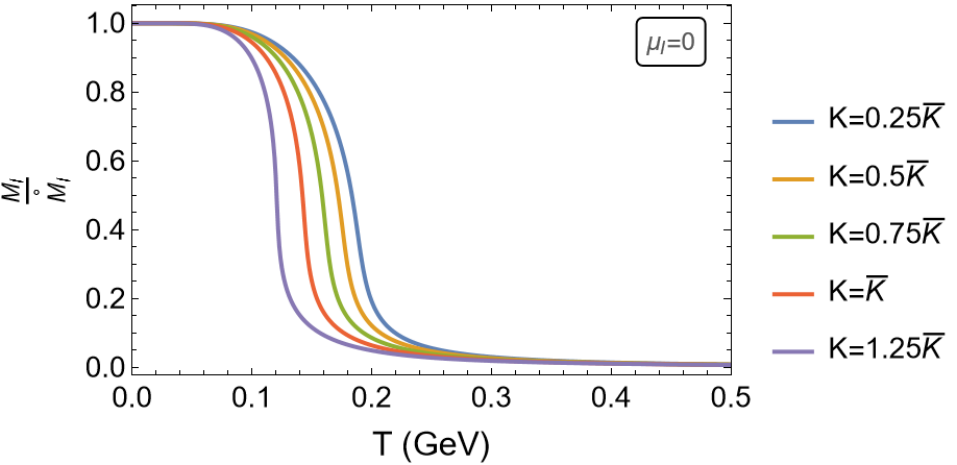}
\vspace{0.4cm}
\includegraphics[width=0.9\linewidth]{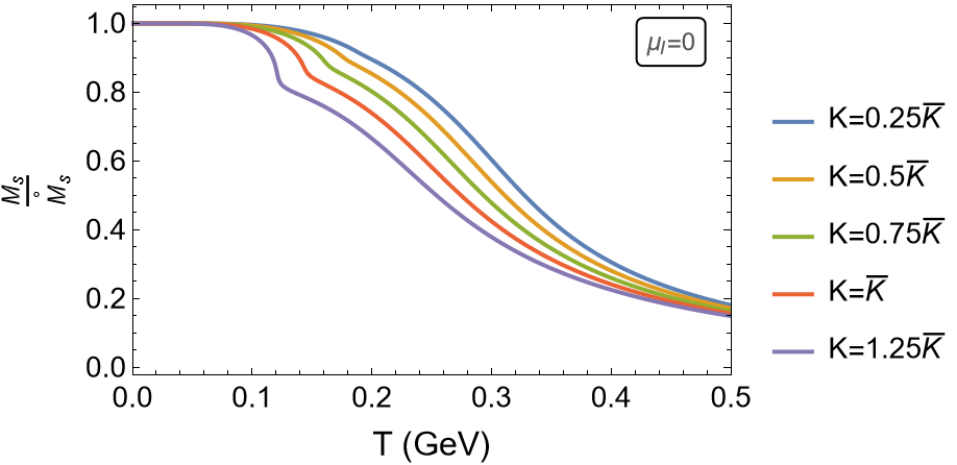}
\caption{$T$ dependencies of the light (top) and strange (bottom) dynamical quark masses, i.e., the respective chiral condensates, with different KMT coupling at $\mu_I=0$. The results are scaled by their own vacuum value.}
\label{fig:quark mass kmt with T}
\end{figure}

We first present $T$ dependencies of the chiral condensates, $M_l$ and $M_s$, at $\mu_I=0$, with several choices of the KMT coupling, in Fig.~\ref{fig:quark mass kmt with T}. The top panel demonstrates the crossover restoration of chiral symmetry for light flavors in hot medium. The strange chiral condensate exhibits the less-pronounced restoration as indicated in the bottom panel, reflecting its large current-quark mass.

The top panel in Fig.~\ref{fig:quark mass kmt with T} also implies that the chiral transition temperature decreases as the anomalous coupling is strengthened, which can be understood by $K$ dependences of the potential. The value of $M_l$ is determined at the minimum point of the potential, which is primarily controlled by a competition between quartic and quadratic coefficients at the mean-field level: $V_{\rm MF} =  (c_{M_l}/2) M_l^2 + (b_{M_l}/4) M_l^4  + \cdots$ with $b_{M_l}=\lambda_2/(6g^4) > 0$ and $c_{M_l}=-m_M^2/g^2-2M_s K/g^3 < 0$. Within the present inputs, both $|c_{M_l}|$ and $b_{M_l}$ decrease with $K$ increased, for which the potential depth at the minimum point gets suppressed. This tendency implies that thermal fluctuations become easier to break chiral condensate, resulting in lower pseudocritical temperature with larger $K$. 

The evident kink in the evolution of $M_s$, which coincides with the pseudocritical temperature of $M_l$, is a direct consequence of the flavor mixing induced by the KMT interaction from the light sector. Thus, a stronger KMT coupling leads to a more pronounced kink structure in the temperature dependence of $M_s$~\cite{Liu:2026cpr,PhysRevD.71.116002,PhysRevD.79.116003}.

\subsection{Pion condensate at finite $\mu_I$}
\label{sec:DeltaAnomaly}

\begin{figure}[htbp]
\centering
\includegraphics[width=0.86\linewidth]{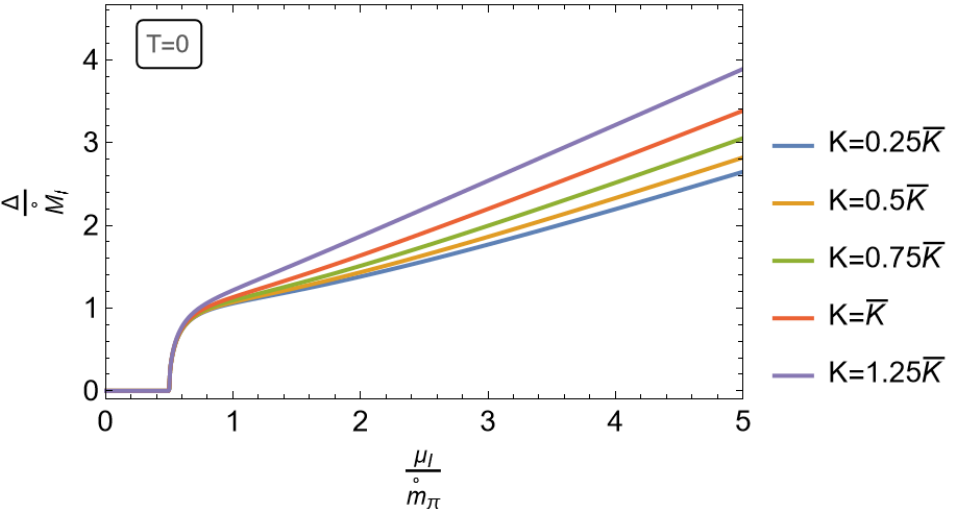}
\vspace{0.4cm}
\includegraphics[width=0.9\linewidth]{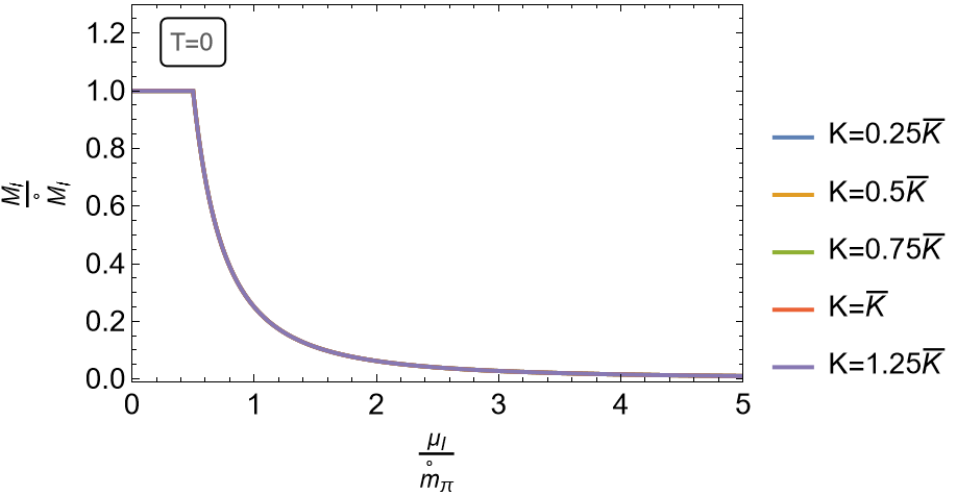}
\vspace{0.4cm}
\includegraphics[width=0.9\linewidth]{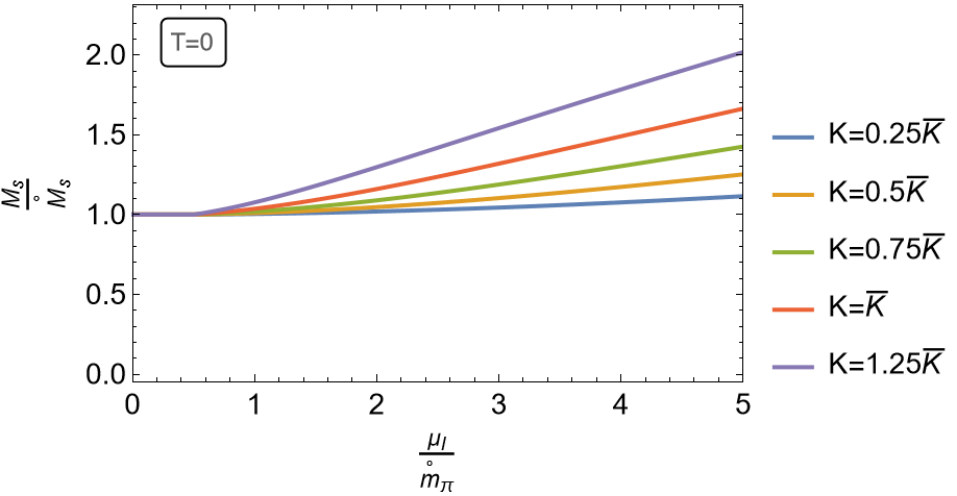}
\caption{$\mu_I$ dependencies of the order parameters, $\Delta$ (top), $M_l$ (middle), and $M_s$ (bottom), with several choices of the KMT coupling $K$ at $T=0$. $\Delta$ and $M_l$ are scaled by the vacuum value $\mathring{M}_{l}$, while $M_s$ is scaled by $\mathring{M}_s$.}
\label{fig:pion condensation kmt}
\end{figure}

Next, we examine $K$ dependencies of the gap $\Delta$, together with $M_l$ and $M_s$, at finite $\mu_I$ with vanishing temperature, to gain insight into the anomaly effects on the pion superfluidity in cold QCD$_I$ medium.

Depicted in Fig.~\ref{fig:pion condensation kmt} are the resultant $\mu_I$ dependencies of $\Delta$, $M_l$ and $M_s$ with several choices of $K$ at $T=0$. The critical isospin chemical potential $\mu_I^c$ is always fixed to be $\m_I^c=\mathring{M}_{\p}/2$, regardless of $K$. As for evolutions of the mean fields in the superfluid phase, first, no variation in $M_l$ is found. This is because $M_l$ in the superfluid phase is exactly expressed by $M_l = (\mathring{M}_l\mathring{M}_\pi^2/4)\mu_I^{-2}$ which does not include any $K$ dependences. Meanwhile, $\Delta$ and $M_s$ have significant $K$ dependences. Within the present refitting scheme, increasing the KMT coupling modifies both the quadratic and quartic coefficients of $\Delta$ in the potential: $V_{\rm MF} = (c/2)\Delta^2 + (b/4)\Delta^4 + \cdots$. More concretely, the quadratic and quartic coefficients, $|c|$ and $b$, become larger and smaller, respectively, with increasing $K$. As a result, since the stationary condition yields $\D=\sqrt{|c|/b}$, one can readily understand the growth of $\Delta$ with $K$ in the superfluid phase. The $K$ dependence of $M_s$ is simply understood by the flavor-mixing structure of the KMT term. As discussed in Appendix~\ref{ap:GapEquation}, the $K$ term induces a driving force from the light sector to enhance $M_s$, and thus, when $\Delta$ gets magnified with $K$ the resultant $M_s$ also becomes large~\cite{Sakai:2025hrj}.

\begin{figure}[htbp]
\centering
\includegraphics[width=0.9\linewidth]{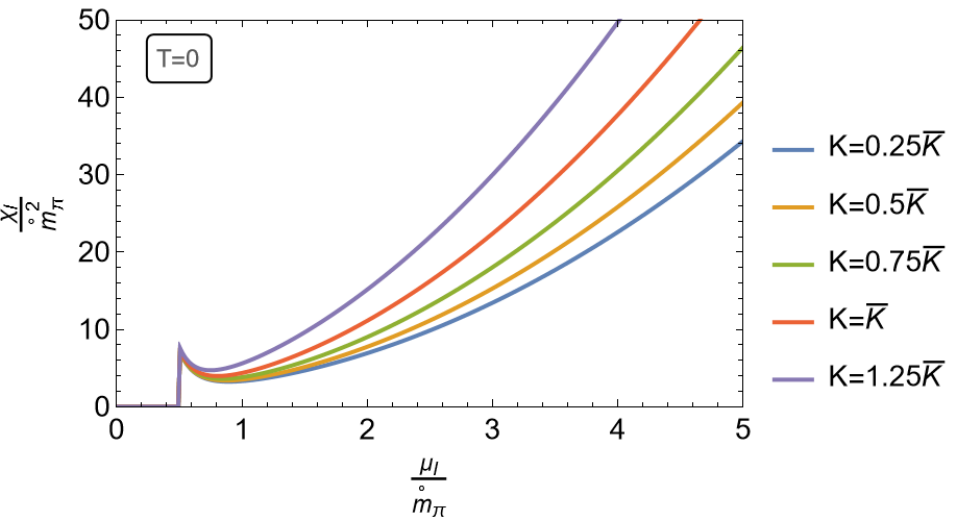}
\vspace{0.4cm}
\includegraphics[width=0.9\linewidth]{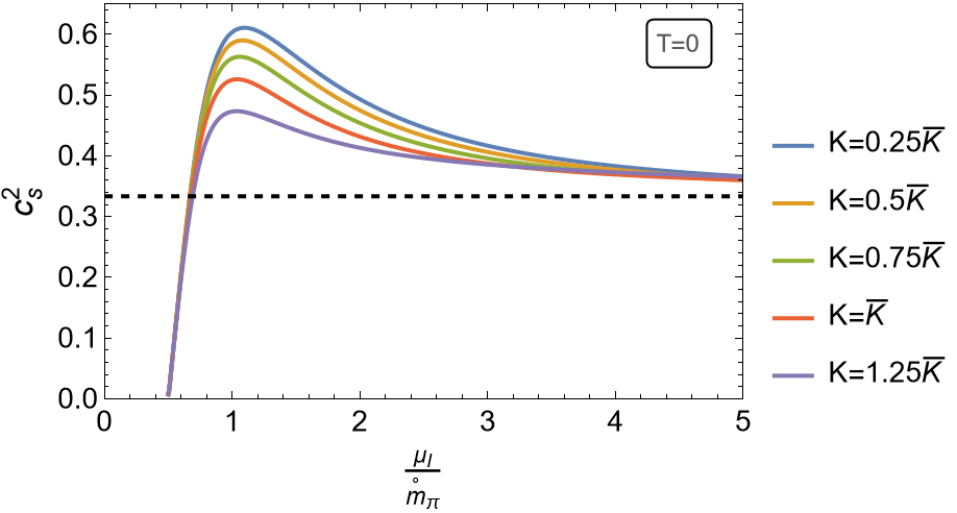}
\includegraphics[width=0.92\linewidth]{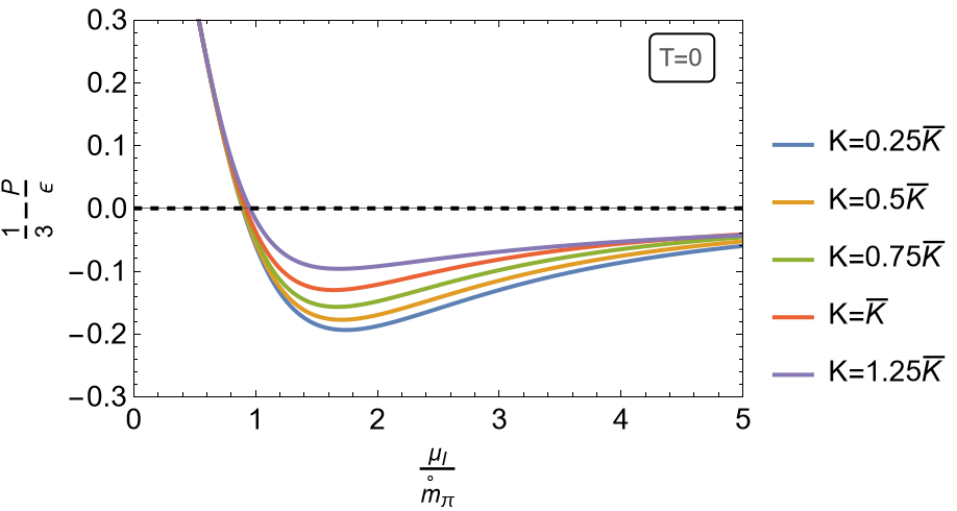}

\vspace{0.4cm}
\caption{$\m_I$ dependencies of the isospin susceptibility (top), sound velocity (middle), and trace anomaly (bottom), with several choices of the KMT coupling $K$ at $T=0$. }
\label{fig:sound speed kmt}
\end{figure}

In Fig.~\ref{fig:sound speed kmt}, we depict $\mu_I$ dependencies of the isospin susceptibility $\chi_I$ (top) and the squared sound velocity (middle), with several values of $K$. This figure indicates that larger $K$ results in enhanced $\chi_I$, i.e., softer EoS in the superfluid phase, and accordingly the peak of $c_s^2$ gets less pronounced. We have also confirmed that larger $K$ leads to smaller $P$ at fixed $\epsilon$, claiming the softening of EoS. Since FRG analysis in the superfluid phase implies an enhancement of $K$ at finite $T$~\cite{Fejos:2026tyd}, it is expected that $c_s^2$ has a rather mild peak above the critical chemical potential.\footnote{Although the FRG analysis done in Ref.~\cite{Fejos:2026tyd} was devoted to dense two-color QCD, we expect a qualitatively similar conclusion in QCD$_I$ medium.} At larger $\mu_I$, $c_s^2$ slowly converges on the conformal limit $\bar{c}_s^2=1/3$ in association with the suppression of instanton effects. 

Since larger $K$ generates the rapid growth of $\Delta$ as demonstrated in Fig.~\ref{fig:pion condensation kmt}, the softening of EoS with larger $K$ is dubbed as the softening with stronger $\Delta$. In the presence of abundant pion superfluidity, the main fraction of isospin medium is carried by the bosonic condensates rather than fermionic quasi-particle excitations near the Fermi surface. Consequently, the system becomes less governed by the Fermi pressure that normally makes matter stiff, but is dominated by the BEC state which can contain more energy with less increased pressure. Thus, medium is more easily compressed, leading to the soft EoS: an increase of $\chi_I$ and a reduction of $c_s^2$.

In Fig.~\ref{fig:sound speed kmt}, we also exhibit the $\mu_I$ dependence of the trace anomaly $\D_{\text{tr}}=\frac{1}{3}-\frac{P}{\e}$ in the bottom panel, the sign of which would include useful information on correlations in cold medium~\cite{Fujimoto:2022ohj}. The figure shows that $\D_{\text{tr}}$ once becomes negative around $\mu_I\sim \mathring{M}_\pi$ and then approaches zero from below. The larger anomaly effect suppresses the depth of negative $\Delta_{\rm tr}$.

\section{Conclusions}
\label{sec:summary}

We have systematically investigated the phase structure, meson mass spectra, EoS and sound velocities in isospin QCD matter, by means of the $N_f=2+1$ quark-meson model that incorporates the Kobayashi-Maskawa-'t Hooft type coupling responsible for the $U(1)_A$ anomaly effects. Within the mean-field approximation at quark one-loop, we find that the transition to the pion superfluid phase, which is of second order at lower temperatures, may become first-order at higher temperatures.

A central focus of this work is the detailed evaluation of the meson mass spectra. Upon entering the superfluid phase, the mass of $\pi^+$ gets exactly zero reflecting the Nambu-Goldstone boson property. Concurrently, the $\pi_0$ mass is found to exhibit strictly linear growth with isospin chemical potential $\mu_I$, which also holds even with thermal fluctuations. Mass degeneracies between chiral partners for light flavors are also seen with the partial restoration of chiral symmetry. 

We also have explored the thermodynamic properties of the isospin medium, which results in that the squared sound velocity generates a peak structure before approaching the conformal limit at lower temperatures. This peak signals the stiffening of the EoS. 
At higher temperatures, it is demonstrated that the possible first-order transition leads to a pronounced discontinuity in the sound velocity.

Moreover, our analysis reveals that the $U(1)_A$ anomaly profoundly impacts medium properties; an enhanced anomaly coupling facilitates pion condensate in dense regime, while it lowers the pseudocritical temperature of chiral restoration in hot medium. The stronger anomaly effects make the sound velocity peak less-pronounced reflecting the softened EoS.

Our findings may provide useful information for future lattice simulations of meson mass spectra and sound velocity from the perspective of symmetry. In what follows, we will provide some comments related to our present work.

Within the present mean-field framework, the pion superfluid transition is found to become first-order at higher temperatures. It is worth noting that, however, fluctuations and other nonperturbative dynamics neglected or incompletely treated at the mean-field level can modify the effective potential near the transition, and consequently affect the critical behavior. In fact, other approaches beyond mean field predict different transition behaviors~\cite{Zhang:2006dn,Xu:2021dki}. Therefore, a more complete treatment of quantum and thermal fluctuations is desirable to investigate structures at higher temperatures properly.

Since our study is based on a mean-field approximation, quantitative variation of the $U(1)_A$ anomaly effects in medium is missing, although we have succeeded in gaining insights into roles of the anomaly effect by changing its coupling strength by hand. In order to study the modification of anomaly effects self-consistently, we are require to adopt a nonperturbative method such as the FRG method~\cite{Fejos:2025oxi,Fejos:2026tyd}. Further investigations based on the FRG are also expected to provide us with more correct treatment of excitations with their fluctuations beyond the present mean-field framework, which may help understand the crossover transition from hadronic to quark matter. We leave such extensive examinations for future study.

It is also instructive to relate the present results to two-color QCD, which offers another sign-problem-free dense QCD matter thanks to the pseudoreality of $SU(2)_c$ group. Although the detailed microscopic dynamics is different from those in QCD$_I$, two-color QCD shares similar phase structures; the lattice results imply the appearance of baryon superfluid phase at low temperature and finite density regime, similarly to the pion superfluidity in QCD$_I$~\cite{Boz:2019enj,Buividovich:2020dks,Astrakhantsev:2020tdl,Iida:2024irv}. Thus, dense two-color QCD serves as a promising testing ground to crosscheck understandings achieved in QCD$_I$ medium, and we leave refined investigation from the quark-meson model in two-color QCD for future study.

\section*{Acknowledgments}
Y.-H.G and X.-G.H. are supported by the National Natural Science Foundation of China (Grants No. 12225502 and No. 12147101), the Natural Science Foundation of Shanghai (Grant No. 23JC1400200), and the National Key Research and Development Program of China (Grant No. 2022YFA1604900). D.S. was supported by Grants-in-Aid for Scientific Research No. 23K03377, No. 23H05439 and No. 25K17386, from Japan Society for the Promotion of Science.

\appendix

\section{Gap equation}
\label{ap:GapEquation}

In the broken phase, the true ground state is determined by the global minimum of the thermodynamic potential. This point necessarily satisfies stationary conditions with respect to $M_l$, $M_s$ and $\Delta$ defined in Eq.~(\ref{Gaps}): $\frac{\partial\Omega}{\partial M_l} = 0$, $\frac{\partial\Omega}{\partial M_s} = 0$ and $\frac{\partial\Omega}{\partial \Delta} = 0$, which read
\begin{eqnarray}
0 &=&  M_l\Bigg[-\frac{m_M^2}{g^2}+\frac{\l_1}{6g^4}(2M_l^2+2\D^2+M_s^2) \nonumber\\
&& +\frac{\l_2}{6g^4}(M_l^2+\D^2)-\frac{2K}{g^3}M_s-\frac{h_l}{gM_l} \nonumber\\
&& +4N_c\sum_{\xi={\rm p,a}}\int_{\vec{k}}\frac{1}{E_l}\frac{E_{l}-\eta_\xi \m_I}{\epsilon_\xi}f(\epsilon_\xi)\Bigg]\ , 
\label{GapSigmal}
\end{eqnarray}
\begin{eqnarray}
0 &=&\frac{M_s}{2} \Bigg[-\frac{m_M^2}{g^2}+\frac{\l_1}{6g^4}(2M_l^2+2\D^2+M_s^2)+\frac{\l_2}{6g^4}M_s^2 \nonumber\\
&& -\frac{2K}{g^3M_s}(M_l^2+\D^2) -\frac{h_s}{gM_s} +8N_c\int_{\vec{k}}\frac{1}{E_{s}}f(E_{s}) \Bigg]\ , \nonumber\\ \label{GapSigmaS}
\end{eqnarray}
and
\begin{eqnarray}
0&=& \D\Bigg[-\frac{4\m_I^2}{g^2}-\frac{m_M^2}{g^2}+\frac{\l_1}{6g^4}(2M_l^2+2\D^2+M_s^2) \nonumber\\
&& +\frac{\l_2}{6g^4}(M_l^2+\D^2) -\frac{2K}{g^3}M_s+4N_c\sum_{\xi={\rm p,a}}\int_{\vec{k}}\frac{f(\epsilon_{\xi})}{\epsilon_{\xi}}\Bigg]\ , \nonumber\\
\label{eq:gap equation} 
\end{eqnarray}
within the no-sea approximation. In these equations we have defined the Fermi-Dirac distribution function
\begin{equation}
f(x)=\frac{1}{\me^{\b x} +1}\ . \label{FDFunction}
\end{equation} 
The $\vec{k}$ dependences of the single-particle excitation energies have been omitted as a shorthand notation:
\begin{eqnarray}
E_l = \sqrt{\vec{k}^2+M_l^2}\ , \ \ E_s = \sqrt{\vec{k}^2+M_s^2}\ , \label{ElEsApp}
\end{eqnarray}
and
\begin{eqnarray}
\epsilon_{\xi} = \sqrt{(E_l-\eta_\xi\mu_I)^2 + \Delta^2}\ \ \ \ \  (\eta_{\rm p/a} = \pm )\ .
\end{eqnarray}

Provided that the quark loop contributions are comparably small, one can confirm 
\begin{eqnarray}
\Delta \sim \sqrt{\frac{24g^2}{\lambda_2}}\mu_I\ , \label{DeltaMuIDep}
\end{eqnarray}
for sufficiently large $\mu_I$, from Eq.~(\ref{eq:gap equation}). The asymptotic behavior of $M_l$ is then determined to be 
\begin{eqnarray}
M_l \sim \frac{g h_l}{4}\mu_I^{-2}  = \frac{\mathring{M}_l \mathring{M}_\pi^2}{4}\mu_I^{-2} \ , \label{MlMuIDep}
\end{eqnarray}
from Eq.~(\ref{GapSigmal}). These expressions imply that $\Delta$ linearly increases with $\mu_I$ while $M_l$ decreases with $\mu_I^{-2}$ independently of the anomaly effects. Hence, the excitation energies $\epsilon_{\rm p}$ and $\epsilon_{\rm a}$ are always sufficiently large and the Boltzmann suppression for the quark fluctuations is significant, for which our assumption that the loop contributions can be neglected is indeed justified in the high density ($\frac{\m_I}{T}\gg1$) region.

The behavior of $M_s$ depends on the anomaly strength $K$. As long as $T$ is rather small, the $s$-quark loop effects are suppressed due to large $M_s$. The $\lambda_1$ contributions are also negligible due to the large $N_c$ suppression. Hence, when $K=0$, the resultant $M_s$ is almost constant against $\mu_I$ from Eq.~(\ref{GapSigmaS}). Meanwhile, with a positive $K$, the anomalous term induces a driving force to enhance $M_s$ in the pion superfluid phase from which $M_l^2+\Delta^2$ increases with $\mu_I$. These structures highlight the role of $U(1)_A$ anomaly effects on $M_s$ in the superfluid phase~\cite{Sakai:2025hrj}.

\section{Meson Mass}
\label{ap:meson mass}

Here, we present tree-level mass and self energies of the mesons which play a central role in evaluating their mass spectra.

\subsection{General properties}
\label{ap:MassGeneral}

In the present paper, we define the meson mass as the so-called pole mass, i.e., an energy $p_0$ at which the corresponding propagator matrix has a pole at rest $\vec{p}=0$.
First, we explain general properties of the propagators.

As illustrated in Table~\ref{tab:meson}, $18$ mesons are grouped into six sectors from the discrete symmetries in the pion superfluid phase, and according the meson propagators would be expressed in a block-diagonalized matrix form. Among them, $\pi_3$ and $\sigma_3$ propagators, $D_{\pi_3}(p)$ and $D_{\sigma_3}(p)$, are always independently obtained. Thus, their inverse propagators are simply expressed by
\begin{eqnarray}
D_{\pi_3}^{-1}(p) &=& p^2-m_{\pi_3\pi_3}^2 -\Pi_{\pi_3\pi_3} (p) \ , \nonumber\\
D_{\sigma_3}^{-1}(p) &=& p^2- m_{\sigma_3\sigma_3}^2 - \Pi_{\sigma_3\sigma_3} (p) \ ,
\end{eqnarray}
where $m_{\pi_3\pi_3}$, $m_{\sigma_3\sigma_3}$ and $\Pi_{\pi_3\pi_3}(p)$, $\Pi_{\sigma_3\sigma_3}(p)$ are the tree-level masses and self energies, respectively. Meanwhile, those of $(\p_1,\p_2,\s_l,\s_s)$, $(\p_l,\p_s,\s_1,\s_2)$ and $(\p_4,\p_5,\s_6,\s_7)$ take the matrix forms of
\begin{eqnarray}
&& D_{\pi_1\pi_2\sigma_l\sigma_s}^{-1} \nonumber\\
&=&  D_0^{-1} -\left(
\begin{array}{cccc}
{\cal M}_{\pi_1\pi_1} & {\cal M}_{\pi_1\pi_2} & {\cal M}_{\pi_1\sigma_l} & {\cal M}_{\pi_1\sigma_s}  \\
{\cal M}_{\pi_1\pi_2} &  {\cal M}_{\pi_2\pi_2} &  {\cal M}_{\pi_2\sigma_l} &  {\cal M}_{\pi_2\sigma_s}  \\
 {\cal M}_{\pi_1\sigma_l} &  {\cal M}_{\pi_2\sigma_l}  &  {\cal M}_{\sigma_l\sigma_l}  &  {\cal M}_{\sigma_l\sigma_s}   \\
 {\cal M}_{\pi_1\sigma_s}  &  {\cal M}_{\pi_2\sigma_s}  &  {\cal M}_{\sigma_l\sigma_s}  &  {\cal M}_{\sigma_s\sigma_s}  \\
\end{array}
\right)\ , \nonumber\\
\end{eqnarray}
\begin{eqnarray}
&& D_{\pi_l\pi_s\sigma_1\sigma_2}^{-1} \nonumber\\
&=&  D_0^{-1} - \left(
\begin{array}{cccc}
 {\cal M}_{\sigma_1\sigma_1}  &  {\cal M}_{\sigma_1\sigma_2} & {\cal M}_{\sigma_1\pi_l} &  {\cal M}_{\sigma_1\pi_s}  \\
 {\cal M}_{\sigma_1\sigma_2}  &  {\cal M}_{\sigma_2\sigma_2}  &  {\cal M}_{\sigma_2\pi_l}  &  {\cal M}_{\sigma_2\pi_s}  \\
 {\cal M}_{\sigma_1\pi_l}  &  {\cal M}_{\sigma_2\pi_l}  &  {\cal M}_{\pi_l\pi_l} & {\cal M}_{\pi_l\pi_s}    \\
 {\cal M}_{\sigma_1\pi_s} &   {\cal M}_{\sigma_2\pi_s}&  {\cal M}_{\pi_l\pi_s} &  {\cal M}_{\pi_s\pi_s} \\
\end{array}
\right)\ , \nonumber\\
\end{eqnarray}
and
\begin{eqnarray}
&& D_{\pi_4\pi_5\sigma_6\sigma_7}^{-1} \nonumber\\
&=& \tilde{D}_0^{-1} - \left(
\begin{array}{cccc}
{\cal M}_{\pi_4\pi_4} & {\cal M}_{\pi_4\pi_5} & {\cal M}_{\pi_4\sigma_6} & {\cal M}_{\pi_4\sigma_7} \\
{\cal M}_{\pi_4\pi_5} & {\cal M}_{\pi_5\pi_5} & {\cal M}_{\pi_5\sigma_6} &  {\cal M}_{\pi_5\sigma_7}  \\
{\cal M}_{\pi_4\sigma_6}  &  {\cal M}_{\pi_5\sigma_6}  &  {\cal M}_{\sigma_6\sigma_6}  &  {\cal M}_{\sigma_6\sigma_7}   \\
 {\cal M}_{\pi_4\sigma_7} &  {\cal M}_{\pi_5\sigma_7} &  {\cal M}_{\sigma_6\sigma_7}  &   {\cal M}_{\sigma_7\sigma_7}  \\
\end{array}
\right)\ , \nonumber\\
\end{eqnarray}
where the ``zero-th'' order propagator inverses $ D_0^{-1} $ and $ \tilde{D}_0^{-1} $ read
\begin{eqnarray}
{D}_0^{-1} = \left(
\begin{array}{cccc}
p^2 & -4{\rm i}\mu_I p_0 & 0 & 0 \\
4{\rm i}\mu_I p_0 & p^2 & 0 & 0 \\
0 & 0 & p^2 & 0 \\
0 & 0 & 0 & p^2 \\
\end{array}
\right)\ ,
\end{eqnarray}
and 
\begin{eqnarray}
\tilde{D}_0^{-1} = \left(
\begin{array}{cccc}
p^2 & -2{\rm i}\mu_I p_0 & 0 & 0 \\
2{\rm i}\mu_I p_0 & p^2 & 0 & 0 \\
0 & 0 & p^2 & 0 \\
0 & 0 & 0 & p^2 \\
\end{array}
\right)\ .
\end{eqnarray}
The mass-function matrix elements take the form of
\begin{eqnarray}
{\cal M}_{\alpha\beta}(p) = m_{\alpha\beta}^2 + \Pi_{\alpha\beta}(p)\ ,
\end{eqnarray}
with the subscripts $\alpha,\beta$ representing the meson fields. We note that the expressions of $(\p_6,\p_7\,\s_4,\s_5)$ are identical to those of $(\p_4,\p_5\,\s_6,\s_7)$ owning to $SU(2)$ isospin symmetry, and have been omitted.

Explicit forms of  the tree-level meson masses $m_{\alpha\beta}^2$ and their self energies generated by the quark one loops $\Pi_{\alpha\beta}(p)$ will be provided in the following subsections.


\subsection{Tree-level meson masses}
\label{ap:TreeLevelMass}


Next, we summarize the tree-level meson masses $m_{\alpha\beta}^2$.

Taking second derivatives of the Lagrangian~(\ref{LMeson}) with respect to the meson fields upon the mean fields~(\ref{Gaps}), one can obtain the following expressions.
\begin{itemize}
\item \underline{Independent modes:}
\end{itemize}
\begin{eqnarray}
m^2_{\p_{3}\p_3}&=&-m^2_M+\frac{\l_1}{6g^2}(2M_l^2+2\D^2+M_s^2) \nonumber\\
&&+\frac{\l_2}{6g^2}(M_l^2+\D^2)-\frac{2K}{g}M_s\ ,  \label{eq:pion3 tree mass} 
\end{eqnarray}
\begin{eqnarray}
m^2_{\s_3\s_3}&=&-m_M^2+\frac{\l_1}{6g^2}(2M_l^2+2\D^2+M_s^2) \nonumber\\
&& +\frac{\l_2}{2g^2}(M_l^2+\D^2)+\frac{2K}{g}M_s\ .
\end{eqnarray}

\begin{itemize}
\item \underline{$(\p_1,\p_2\,\s_l,\s_s)$ group:}
\end{itemize}
\begin{eqnarray}
m^2_{\p_1\p_1}&=& -m^2_M+\frac{\l_1}{6g^2}(2M_l^2+6\D^2+M_s^2) \nonumber\\
&& + \frac{\l_2}{6g^2}(M_l^2+3\D^2)-\frac{2K}{g}M_s-4\m_I^2\ , 
\end{eqnarray}
\begin{eqnarray}
m^2_{\p_{2}\p_2}&=& -m^2_M+\frac{\l_1}{6g^2}(2M_l^2+2\D^2+M_s^2) \nonumber\\
&&+ \frac{\l_2}{6g^2}(M_l^2+\D^2)-\frac{2K}{g}M_s-4\m_I^2\ ,  \label{MPi2Tree} 
\end{eqnarray}
\begin{eqnarray}
m^2_{\s_l\s_l}&=& -m^2_M+\frac{\l_1}{6g^2}(2M_l^2+6\D^2+M_s^2) \nonumber\\
&&+ \frac{\l_2}{6g^2}(3M_l^2+\D^2)-\frac{2K}{g}M_s\ , 
\end{eqnarray}
\begin{eqnarray}
m^2_{\s_s\s_s}&=& -m^2_M+\frac{\l_1}{6g^2}(2M_l^2+2\D^2+3M_s^2) \nonumber\\
&&+ \frac{\l_2}{2g^2}M_s^2\ , 
\end{eqnarray}
\begin{eqnarray}
m^2_{\s_l\s_s}&=& \frac{-2\sqrt{2}K}{g}M_l+\frac{\sqrt{2}\l_1}{3g^2}M_lM_s\ , 
\end{eqnarray}
\begin{eqnarray}
m^2_{\p_1\s_l}&=& \frac{2\l_1}{3g^2}\D M_l+\frac{\l_2}{3g^2}\D M_l\ , 
\end{eqnarray}
\begin{eqnarray}
m^2_{\p_1\s_s}&=& \frac{-2\sqrt{2}K}{g}\D+\frac{\sqrt{2}\l_1}{3g^2}\D M_s\ .
\end{eqnarray}

\begin{itemize}
\item \underline{$(\p_l,\p_s\,\s_1,\s_2)$ group:}
\end{itemize}
\begin{eqnarray}
m^2_{\p_l\p_l}&=& -m_M^2+\frac{\l_1}{6g^2}(2M_l^2+2\D^2+M_s^2) \nonumber\\
&& + \frac{\l_2}{6g^2}(M_l^2+3\D^2)+\frac{2K}{g}M_s \ ,   \label{MEtaL}
\end{eqnarray}
\begin{eqnarray}
m^2_{\p_s\p_s}&=& -m_M^2+\frac{\l_1}{6g^2}(2M_l^2+2\D^2+M_s^2) \nonumber\\
&&+ \frac{\l_2}{6g^2}M_s^2\ ,  \label{MEtaS}
\end{eqnarray}
\begin{eqnarray}
m^2_{\p_l\p_s}&=& \frac{2\sqrt{2}K}{g}M_l\ ,  \label{MEtaLS}
\end{eqnarray}
\begin{eqnarray}
m^2_{\s_1\s_1}&=& -m_M^2+\frac{\l_1}{6g^2}(2M_l^2+2\D^2+M_s^2) \nonumber\\
&&+ \frac{\l_2}{6g^2}(3M_l^2+\D^2)+\frac{2K}{g}M_s-4\m_I^2\ , 
\end{eqnarray}
\begin{eqnarray}
m^2_{\s_2\s_2}&=& -m_M^2+\frac{\l_1}{6g^2}(2M_l^2+2\D^2+M_s^2) \nonumber\\
&&+ \frac{\l_2}{2g^2}(M_l^2+\D^2)+\frac{2K}{g}M_s-4\m_I^2\ , 
\end{eqnarray}
\begin{eqnarray}
m^2_{\p_l\s_1}&=& \frac{\l_2}{3g^2}\D M_l\ , 
\end{eqnarray}
\begin{eqnarray}
m^2_{\p_s\s_1}&=& -\frac{2\sqrt{2}K}{g}\D\  . 
\end{eqnarray}

\begin{itemize}
\item \underline{$(\p_4,\p_5\,\s_6,\s_7)$ group:}
\end{itemize}
\begin{eqnarray}
m^2_{\p_4\p_4} &=& -m^2_M+\frac{\l_1}{6g^2}(2M_l^2+2\D^2+M_s^2) \nonumber\\
&& + \frac{\l_2}{6g^2}(M_l^2+M_s^2-M_lM_s+\D^2) \nonumber\\
&&-\frac{2K}{g}M_l-\m_I^2\ , \\
m^2_{\s_6\s_6}&=& -m^2_M+\frac{\l_1}{6g^2}(2M_l^2+2\D^2+M_s^2) \nonumber\\
&&+ \frac{\l_2}{6g^2}(M_l^2+M_s^2+M_lM_s+\D^2) \nonumber\\
&& +\frac{2K}{g}M_l-\m_I^2\ , \\
m^2_{\s_6\p_4}&=& m^2_{\s_7\p_5}=\frac{\l_2}{6g^2}\D M_s+\frac{2K}{g}\D\ , 
\end{eqnarray}
\begin{eqnarray}
m^2_{\p_5\p_5} = m^2_{\p_4\p_4}\ , \ \ m^2_{\s_7\s_7} = m^2_{\s_6\s_6}\ .
\end{eqnarray} 

When $\mu_I$ is sufficiently large, the light-quark dynamical mass vanishes, $M_l\to0$, and thus $m^2_{\p_4\p_4} = m^2_{\p_5\p_5} = m^2_{\s_6\s_6} = m^2_{\s_7\s_7}$ further holds in $(\p_4,\p_5\,\s_6,\s_7)$ group. These symmetric mass matrix yields $M_{K_+}=M_{K_-}$ and $M_{\kappa_+} = M_{\kappa_-}$ as numerically confirmed at $\mu_I\sim 0.4$ GeV in Fig.~\ref{fig:meson mass T=0}. We note that the anomaly effects vanish for the $K_\pm$ and $\kappa_\pm$ in this limit, since the $K$ term is accompanied by $M_l$ for those open-strange mesons.

In this subsection, we have only shown nonzero tree-level masses. The remaining ones are vanishing.

\subsection{Self energies}
\label{ap:SelfEnergies}

Finally, here we exhibit the meson self energies $\Pi_{\alpha\beta}$ generated by the quark one loops via the Yukawa coupling.

Making use of the quark propagator matrix~(\ref{SMatix}) with Eqs.~(\ref{SlightElement}) and~(\ref{Sstrange}), the self energies within the no-sea approximation at the rest frame $\vec{p}={0}$ are calculated as follows.
 
\begin{widetext}
\begin{itemize}
\item \underline{Independent modes:}
\end{itemize}
\begin{eqnarray}
\P_{\p_3\p_3} (p) &=&-\mi g^2\int_k\mtr[ \l_3\mi\g^5S_q(k+p)\l_3\mi\g^5S_q(k)] \nonumber\\
&\overset{\vec{p}\to{0}}{=}& 
 -4g^2N_c\sum_{\zeta=\pm} \int_{\vec{k}} \left(1-\zeta\frac{E_l^2-\mu_I^2+\Delta^2}{\epsilon_{\rm p}\epsilon_{\rm a}} \right)\frac{\epsilon_{\rm p}-\zeta\epsilon_{\rm a}}{p_0^2-(\epsilon_{\rm p}-\zeta\epsilon_{\rm a})^2}  \Big(f\big(\epsilon_{\rm p}\big)-\zeta f\big(\epsilon_{\rm a}\big)\Big)  \ , \label{Pi3Pi3}
\end{eqnarray}
and
\begin{eqnarray}
\P_{\s_3\s_3} (p) &=& -\mi g^2\int_k\mtr[\l_3S_q(k+p)\l_3S_q(k)] \nonumber\\
&\overset{\vec{p}\to{0}}{=}&
 -4g^2N_c \sum_{\zeta=\pm}\int_{\vec{k}}  \left(1-\zeta\frac{E_l^2-\mu_I^2-\Delta^2}{\epsilon_{\rm p}\epsilon_{\rm a}} \right) \frac{\vec{k}^2}{E_l^2} \frac{\epsilon_{\rm p}-\zeta\epsilon_{\rm a}}{p_0^2-(\epsilon_{\rm p} -\zeta\epsilon_{\rm a})^2}  \Big( f\big(\epsilon_{\rm p}\big)-\zeta f\big(\epsilon_{\rm a}\big)\Big) \ .
\end{eqnarray}

\begin{itemize}
\item \underline{$(\p_1,\p_2,\s_l,\s_s)$ group:}
\end{itemize}
\begin{eqnarray}
\P_{\p_1\p_1}(p) =-\mi g^2\int_k\mtr[ \l_1\mi\g^5S_q(k+p)\l_1\mi\g^5S_q(k)] \overset{\vec{p}\to{0}}{=}
-16g^2N_c\sum_{\xi={\rm p,a}}\int_{\vec{k}} \frac{(E_l-\eta_\xi \mu_I)^2}{\epsilon_\xi} \frac{1}{p_0^2-4\epsilon_{\xi}^2} f\big(\epsilon_\xi\big)  \ ,
\end{eqnarray}

\begin{eqnarray}
\P_{\p_2\p_2}(p) = -\mi g^2\int_k\mtr[ \l_2\mi \g^5S_q(k+p)\l_2\mi\g^5S_q(k)] \overset{\vec{p}\to{0}}{=}  
-16g^2N_c \sum_{\xi={\rm p,a}}\int_{\vec{k}}  \frac{\epsilon_\xi}{p_0^2-4\epsilon_\xi^2}  f\big(\epsilon_\xi\big)\ , \label{SEPi2} 
\end{eqnarray}

\begin{eqnarray}
\P_{\p_1\p_2} (p)=-\mi g^2\int_k\mtr[\l_1\mi\g^5S_q(k+p)\l_2\mi\g^5S_q(k)] 
\overset{\vec{p}\to{0}}{=}
8 \mi g^2N_c\sum_{\xi={\rm p,a}}\int_{\vec{k}} \eta_\xi\frac{E_l-\eta_\xi\mu_I}{\epsilon_\xi} \frac{p_0}{p_0^2- 4\epsilon_\xi^2}  f\big(\epsilon_{\xi}\big) \ ,
\end{eqnarray}

\begin{eqnarray}
\P_{\s_l\s_l}(p) &=& -\mi g^2\int_k\mtr[\l_lS_q(k+p)\l_lS_q(k)] \nonumber\\
&\overset{\vec{p}\to{0}}{=}& 
- 8g^2N_c\int_{\vec{k}}  \Bigg\{\sum_{\xi={\rm p,a}}\frac{M_l^2}{E_l^2}\frac{\Delta^2}{\epsilon_\xi}\frac{2}{p_0^2- 4\epsilon_{\xi}^2 }f\big(\epsilon_{\xi} \big)  \nonumber\\
&& +\sum_{\zeta=\pm} \frac{\vec{k}^2}{2E_l^2}\left(1-\zeta\frac{E_l^2-\mu_I^2+\Delta^2}{\epsilon_{\rm p}\epsilon_{\rm a}} \right)\frac{(\epsilon_{\rm p}-\zeta\epsilon_{\rm a})}{p_0^2-(\epsilon_{\rm p}-\zeta\epsilon_{\rm a})^2} \Big( f\big(\epsilon_{\rm p} \big)-\zeta f\big(\epsilon_{\rm a}\big)\Big) \Bigg\} \ ,
\end{eqnarray}

\begin{eqnarray}
\P_{\s_s\s_s}(p) =-\mi g^2\int_k\mtr[\l_sS_q(k+p)\l_sS_q(k)] \overset{\vec{p}\to{0}}{=} -32g^2N_c\int_{\vec{k}}  \frac{\vec{k}^2}{E_s^2}  \frac{E_s}{q_0^2- 4E_s^2}f\big(E_s\big)\ ,
\end{eqnarray}

\begin{eqnarray}
\P_{\s_l\p_1} (p) = -\mi g^2\int_{k}\mtr[\l_lS_q(k+p)\l_1\mi\g^5S_q(k)] 
\overset{\vec{p}\to{0}}{=} 
16g^2N_c \sum_{\xi={\rm p,a}}\int_{\vec{k}} \frac{M_l}{E_l}\frac{(E_l-\eta_\xi\mu_I)\Delta}{\epsilon_\xi}\frac{1}{p_0^2- 4\epsilon_{\xi}^2}  f\big(\epsilon_{\xi}\big)\ ,
\end{eqnarray}
and
\begin{eqnarray}
\P_{\s_l\p_2} (p) = -\mi g^2\int_k\mtr[\l_lS_q(k+p)\l_2\mi\g^5S_q(k)] \overset{\vec{p}\to{0}}{=}
-8\mi g^2N_c\sum_{\xi={\rm p,a}}\int_{\vec{k}}  \eta_\xi \frac{M_l}{E_l}\frac{\Delta}{\epsilon_\xi}\frac{p_0}{p_0^2- 4\epsilon_{\xi}^2} f\big(\epsilon_{\xi}\big) \ .
\end{eqnarray}

\begin{itemize}
\item \underline{$(\p_l,\p_s,\s_1,\s_2)$ group:}
\end{itemize}

\begin{eqnarray}
\P_{\s_1\s_1} (p) &=& -\mi g^2\int_k\mtr[\l_1S_q(k+p)\l_1S_q(k)] \nonumber\\
&\overset{\vec{p}\to{0}}{=}& 
- 8g^2N_c\int_{\vec{k}}  \Bigg\{\sum_{\xi={\rm p,a}}\frac{\vec{k}^2}{E_l^2}\frac{2\epsilon_\xi}{p_0^2 - 4\epsilon_{\xi}^2}  f\big(\epsilon_{\xi}\big) \nonumber\\
&& +\sum_{\zeta=\pm} \frac{M_l^2}{2E_l^2}\left(1+\zeta \frac{E_l^2-\mu_I^2-\Delta^2}{\epsilon_{\rm p}\epsilon_{\rm a}}\right) \frac{(\epsilon_{\rm p}-\zeta\epsilon_{\rm a})}{p_0^2-(\epsilon_{\rm p}-\zeta\epsilon_{\rm a})^2}  \Big( f\big(\epsilon_{\rm p}\big)-\zeta f\big(\epsilon_{\rm a}\big)\Big) \Bigg\}\ ,
\end{eqnarray}

\begin{eqnarray}
\P_{\s_2\s_2} (p) &=& -\mi g^2\int_k\mtr[\l_1S_q(k+p)\l_1S_q(k)] \nonumber\\
&\overset{\vec{p}\to{0}}{=}&
-8g^2N_c\int_{\vec{k}}  \Bigg\{\sum_{\xi={\rm p,a}}\frac{\vec{k}^2}{E_l^2}\frac{(E_l-\eta_\xi\mu_I)^2}{\epsilon_\xi}\frac{2}{p_0^2 - 4\epsilon_{\xi}^2} f_F\big(\epsilon_{\xi}\big)  + \sum_{\zeta=\pm }\frac{M_l^2}{E_l^2} \frac{(\epsilon_{\rm p}-\zeta\epsilon_{\rm a})}{p_0^2-(\epsilon_{\rm p}-\zeta\epsilon_{\rm a})^2}\Big( f\big(\epsilon_{\rm p}\big)-\zeta f\big(\epsilon_{\rm a}\big)\Big)   \Bigg\} \ , \nonumber\\
\end{eqnarray}

\begin{eqnarray}
\P_{\s_1\s_2} (p)&=&-\mi g^2\int_k\mtr[\l_1S_q(k+p)\l_2S_q(k)] \nonumber\\
&\overset{\vec{p}\to{0}}{=}& 
 8\mi g^2N_c\int_{\vec{k}}  \Bigg\{\sum_{\xi={\rm p,a}}\eta_\xi \frac{\vec{k}^2}{E_l^2}\frac{E_l-\eta_\xi\mu_I}{\epsilon_\xi}\frac{p_0}{p_0^2- 4\epsilon_{\xi}^2} f\big(\epsilon_{\xi}\big) \nonumber\\
&& + \frac{M_l^2}{2E_l^2}\left(\frac{E_l-\mu_I}{\epsilon_{\rm p}}+\zeta\frac{E_l+\mu_I}{\epsilon_{\rm a}}\right) \frac{p_0}{p_0^2-(\epsilon_{\rm p}-\zeta\epsilon_{\rm a})^2} \Big( f\big(\epsilon_{\rm p}\big)-\zeta f\big(\epsilon_{\rm a}\big)\Big) \Bigg\}\ ,
\end{eqnarray}

\begin{eqnarray}
\P_{\p_l\p_l} (p) &=&-\mi g^2\int_k\mtr[\l_l \mi \g^5S_q(k+p)\l_l \mi \g^5S_q(k)] \nonumber\\
&\overset{\vec{p}\to{0}}{=}& 
-4g^2N_c \sum_{\zeta=\pm}\int_{\vec{k}} \left(1-\zeta \frac{E_l^2-\mu_I^2-\Delta^2}{\epsilon_{\rm p}\epsilon_{\rm a}}\right) \frac{\epsilon_{\rm p} -\zeta \epsilon_{\rm a}}{p_0^2-(\epsilon_{\rm p}-\zeta\epsilon_{\rm a})^2}  \Big( f\big(\epsilon_{\rm p}\big)- \zeta f\big(\epsilon_{\rm a}\big)\Big) \ ,
\end{eqnarray}

\begin{eqnarray}
\P_{\p_s\p_s}(p)  = -\mi g^2\int_k\mtr[\l_s\mi\g^5S_q(k+p)\l_s\mi\g^5S_q(k)] \overset{\vec{p}\to{0}}{=} -32g^2N_c\int_{\vec{k}} \frac{E_s}{p_0^2-4E_s^2} f(E_s)\ ,
\end{eqnarray}

\begin{eqnarray}
\P_{\s_1\p_l} (p) &=& -\mi g^2\int_k\mtr[\l_1S_q(k+p)\l_l\mi\g^5S_q(k)] \nonumber\\
&\overset{\vec{p}\to{0}}{=}&
- 8g^2N_c \sum_{\zeta=\pm}\int_{\vec{k}}  \frac{\zeta M_l\Delta}{\epsilon_{\rm p}\epsilon_{\rm a}} \frac{(\epsilon_{\rm p} -\zeta \epsilon_{\rm a})}{p_0^2-(\epsilon_{\rm p}-\zeta\epsilon_{\rm a})^2}\Big( f\big(\epsilon_{\rm p}\big)- \zeta f\big(\epsilon_{\rm a}\big)\Big) \ ,
\end{eqnarray}
and

\begin{eqnarray}
\P_{\s_2\p_l} (p) &=& -\mi g^2\int_k\mtr[\l_2S_q(k+p)\l_l\mi\g^5S_q(k)] \nonumber\\
&\overset{\vec{p}\to{0}}{=}& 
-4\mi g^2N_c \sum_{\zeta=\pm}\int_{\vec{k}}  \frac{M_l}{E_l} \left(\frac{\Delta}{\epsilon_{\rm p}} +\zeta \frac{\Delta}{\epsilon_{\rm a}}\right) \frac{p_0}{p_0^2-(\epsilon_{\rm p}-\zeta\epsilon_{\rm a})^2} \Big( f\big(\epsilon_{\rm p}\big)- \zeta f\big(\epsilon_{\rm a}\big)\Big) \ .
\end{eqnarray}

\begin{itemize}
\item \underline{$(\p_6,\p_7,\s_4,\s_5)$ group:}
\end{itemize}

\begin{eqnarray}
\P_{\s_4\s_4} (p) &=& -\mi g^2\int_k\mtr [\l_5S_q(k+p)\l_5S_q(k)] \nonumber\\
&\overset{\vec{p}\to{0}}{=}& 
-2g^2N_c\int_{\vec{k}}\sum_{\xi={\rm p,a}}\Bigg\{ \sum_{\zeta=\pm} \left(1-\zeta\frac{E_l-\eta_\xi\mu_I}{\epsilon_\xi}\frac{\vec{k}^2 - M_{s}M_l}{E_lE_s} \right) \frac{\epsilon_\xi-\zeta E_s}{p_0^2-(\epsilon_{\xi} - \zeta E_s )^2}  \Big(f\big(\epsilon_{\xi}\big) - \zeta f\big(E_s\big)  \Big) \Bigg\} \ , \nonumber\\
\end{eqnarray} 

\begin{eqnarray}
\P_{\s_4\s_5}(p)  &=& -\mi g^2\int_k\mtr[\l_4S_q(k+p)\l_5S_q(k)] \nonumber\\
&\overset{\vec{p}\to{0}}{=}&
 2ig^2N_c\int_{\vec{k}}\sum_{\xi={\rm p,a}} \Bigg\{ \eta_\xi \sum_{\zeta=\pm}   \left(\frac{E_l-\eta_\xi\mu_I}{\epsilon_\xi}-\zeta\frac{\vec{k}^2-M_{s}M_l}{E_lE_s} \right) \frac{p_0}{p_0^2-(\epsilon_{\xi} -\zeta E_s )^2}  \Big(f\big(\epsilon_{\xi}\big) - \zeta f\big(E_s\big)  \Big)  \Bigg\} \ ,  \nonumber\\
\end{eqnarray}
\begin{eqnarray}
\P_{\p_6\p_6} (p) &=& -\mi g^2\int_k\mtr[\mi\g^5\l_6S_q(k+p)\mi\g^5\l_6S_q(k)] \nonumber\\
&\overset{\vec{p}\to{0}}{=}&
 -2g^2N_c\int_{\vec{k}}\sum_{\xi={\rm p,a}} \Bigg\{  \sum_{\zeta=\pm} \left(1-\zeta\frac{E_l-\eta_\xi\mu_I}{\epsilon_\xi}\frac{\vec{k}^2 + M_{s}M_l}{E_lE_s} \right) \frac{\epsilon_\xi- \zeta E_s}{p_0^2-(\epsilon_{\xi} - \zeta E_s )^2}  \Big(f\big(\epsilon_{\xi}\big) - \zeta f\big(E_s\big)  \Big) \Bigg\}\ , \nonumber\\
\end{eqnarray}

\begin{eqnarray}
\P_{\p_6\p_7}(p) &=& -\mi g^2\int_k\mtr[\mi\g^5\l_6S_q(k+p)\mi\g^5\l_7S_q(k)] \nonumber\\
&\overset{\vec{p}\to{0}}{=}& 
 -2\mi g^2N_c \int_{\vec{k}}\sum_{\xi={\rm p,a}}  \Bigg\{\eta_\xi  \sum_{\zeta=\pm} \left(\frac{E_l-\eta_\xi\mu_I}{\epsilon_\xi}-\zeta \frac{\vec{k}^2+M_{s}M_l}{E_lE_s} \right) \frac{p_0}{p_0^2-(\epsilon_{\xi}-\zeta E_s )^2}  \Big(f\big(\epsilon_{\xi}\big) - \zeta f\big(E_s\big)  \Big)  \Bigg\} \ , \nonumber\\
\end{eqnarray}

\begin{eqnarray}
\P_{\s_4\p_6} (p) &=&-\mi g^2\int_k\mtr[\l_4S_q(k+p) \l_6\mi \g^5S_q(k)] \nonumber\\
&\overset{\vec{p}\to{0}}{=}&
- 2g^2N_c \int_{\vec{k}} \sum_{\xi={\rm p,a}} \Bigg\{ \sum_{\zeta=\pm}  \frac{\zeta M_s \Delta }{E_s\epsilon_{\xi}} \frac{\epsilon_\xi- \zeta E_s}{q_0^2-(\epsilon_{\xi} - \zeta E_s )^2}  \Big(f\big(\epsilon_{\xi}\big) - \zeta f\big(E_s\big)  \Big) \Bigg\}\ ,
\end{eqnarray}

and
\begin{eqnarray}
\P_{\s_4\p_7} (p) &=& -\mi g^2\int_k\mtr[\l_4S_q(k+p) \l_7\mi \g^5S_q(k)] \nonumber\\
&\overset{\vec{p}\to{0}}{=}&  
2 \mi g^2N_c \int_{\vec{k}} \sum_{\xi={\rm p,a}}  \Bigg\{ \eta_\xi \sum_{\zeta=\pm} \frac{M_l\Delta}{E_l\epsilon_\xi}\frac{p_0}{p_0^2-(\epsilon_{\xi} -\zeta E_s )^2}  \Big(f\big(\epsilon_{\xi}\big) - \zeta f\big(E_s\big)  \Big) \Bigg\}\ .
\end{eqnarray}
\end{widetext}

In these equations, the symbol ``Tr'' stands for the trace operator for the Dirac, color and flavor spaces, and $\int_k$ represents the Matsubara summations as well as the three-dimensional momentum integrals. The $\vec{k}$ dependences of the excitation energies $E_l$, $E_s$, $\epsilon_{\rm p}$ and $\epsilon_{\rm a}$ must be understood implicitly as in Eq.~(\ref{ElEsApp}). The Matsubara summations have been performed by means of
\begin{eqnarray}
 T\sum_m\frac{1}{(i\omega_m+i\bar{\omega}_n-\epsilon_1)(i\omega_m-\epsilon_2)} = \frac{f(\epsilon_2)-f(\epsilon_1)}{i\bar{\omega}_n-\epsilon_1+\epsilon_2} \ , \nonumber\\
\end{eqnarray}
with fermionic and bosonic Matsubara frequencies within the imaginary-time formalism: $\omega_m=(2m+1)\pi T$ and $\bar{\omega}_n=2n\pi T$ ($m,n \in {\mathbb Z}$). We note that the self-energy matrix satisfies $(\Pi_{\beta\alpha})^* = \Pi_{\alpha\beta}$ owing to its hermiticity.

\section{Ginzburg-Landau analysis}
\label{ap:Laudau theory}

Here, we proceed with the Ginzburg-Landau (GL) analysis to gain a better understanding into the transition to the pion superfluid phase.

The GL framework is based on an expansion with respect to the order parameter $\Delta$. When the external source of $\Delta$ is absent, the
thermodynamic potential is invariant under $\Delta\rightarrow-\Delta$, such that only
even powers of $\Delta$ appear. Typically it is sufficient to expand up to the fourth order:
\begin{equation}
\O=\frac{c}{2}\D^2+\frac{b}{4}\D^4\ .
\label{eq: quartic potentential}
\end{equation}
In this case the coefficient $b$ must be positive to make system stable. When $c>0$, the effective potential just has one minimum at $\Delta=0$ and there is no symmetry broken. When $c<0$, the free energy has a nontrivial minimum at $\Delta\neq0$ and there exists a symmetry breaking. However, $b$ could be negative, and in this case it is necessary to expand the potential up to the sixth order
\begin{equation}
\O=\frac{c}{2}\D^2+\frac{b}{4}\D^4+\frac{a}{6}\D^6 \ \ \ \ \  ({\rm with}\ a>0)\ .
\end{equation}
In principle, the function can have two local minimum when we restrict ourselves for $\Delta\geq0$. The first order phase transition will occur when the global minimum jumps from one local minimum to another.

Let us consider a thermodynamic potential with multiple order parameters,
\begin{equation}
    \Omega(\Delta^2,\boldsymbol{\Phi})\ ,
\end{equation}
where we have used a fact that $\Omega$ is a function of $\Delta^2$ and the other order parameters
$\boldsymbol{\Phi}=(\phi_1,\phi_2,\phi_3,\ldots)^T$. In the normal phase, we choose the reference point
$\Delta=0$ and $\boldsymbol{\Phi}=\boldsymbol{\Phi}_0$, where
$\boldsymbol{\Phi}_0$ is determined by the gap equations
\begin{equation}
\left.\frac{\partial\Omega}{\partial{\bm \Phi}}\right|_{\D=0,\boldsymbol{\F}=\boldsymbol{\F}_0}=0\ . \label{GapEqDelta0}
\end{equation}
In the vicinity of phase-transition point, ${\bm \Phi}$ can be approximated by a power series of $\Delta^2$:
\begin{equation}
{\bm \Phi}(\Delta^2)={\bm \Phi}_{0}+{\bm \g}\Delta^2+O(\Delta^4)\ .
\end{equation}
That is, $\delta{\bm \Phi}$ defined by [$\delta\boldsymbol{\Phi}=(\delta\phi_1,\delta\phi_2,\delta\phi_3,\ldots)^T$]
\begin{equation}
    \delta\boldsymbol{\Phi}
    \equiv
    \boldsymbol{\Phi}-\boldsymbol{\Phi}_0
    =\boldsymbol{\g}\Delta^2+O(\Delta^4)
\end{equation}
starts from ${\cal O}(\Delta^2)$. Then, expanding the gap equations for $\boldsymbol{\Phi}$: $\frac{\partial\Omega}{\partial\boldsymbol{\Phi}}=0$, with respect to $\Delta^2$ around $\Delta=0$ gives
\begin{eqnarray}
 0 
=  \left.\frac{\partial\Omega}{\partial{\bm \Phi}}\right|_{\Delta=0,{\bm \Phi} = {\bm \Phi}_0} + \left({\bm H}{\bm \gamma}+{\bm v}\right)\Delta^2 + {\cal O}(\Delta^4) \ , \label{GapEqExpand}
\end{eqnarray}
where
\begin{equation}
    H_{ij}
    \equiv
    \left.
    \frac{\partial^2\Omega}
    {\partial\phi_i\partial\phi_j}
    \right|_{\D=0,\boldsymbol{\F}=\boldsymbol{\F}_0} \label{Hessian}
\end{equation}
is the Hessian matrix in the $\boldsymbol{\Phi}$ subspace, and
\begin{equation}
    v_i
    \equiv
    \left.
    \frac{\partial^2\Omega}
    {\partial\phi_i\partial\Delta^2}
    \right|_{\D=0,\boldsymbol{\F}=\boldsymbol{\F}_0}\ . \label{vHv}
\end{equation}
The ${\cal O}(\Delta^0)$ part in Eq.~(\ref{GapEqExpand}) vanishes as required by Eq.~(\ref{GapEqDelta0}). Similarly, from the coefficients of $\Delta^2$ one can get
\begin{equation}
    \boldsymbol{\g}=-{\bm H}^{-1}\boldsymbol{v}\ . \label{GammaExp}
\end{equation}

Expanding the thermodynamic potential $\Omega$ in the same way, we obtain
\begin{eqnarray}
\Omega &\approx& \Omega(0,{\bm \Phi_0}) + \frac{\partial\Omega}{\partial\phi_i}\delta\phi_i + \frac{\partial\Omega}{\partial\Delta^2}\Delta^2  + \frac{1}{2}\Bigg(\frac{\partial^2\Omega}{\partial\phi_i\partial\phi_j}\delta\phi_i\delta\phi_j \nonumber\\
&& + 2\frac{\partial^2\Omega}{\partial\phi_i\partial\Delta^2}\delta\phi_i\Delta^2 + \frac{\partial^2\Omega}{\partial(\Delta^2)^2}(\Delta^2)^2\Bigg) \nonumber\\
&\approx&  \Omega(0,{\bm \Phi_0}) +  \frac{\partial\Omega}{\partial\Delta^2}\Delta^2 + \frac{1}{2}\Bigg(  \frac{\partial^2\Omega}{\partial(\Delta^2)^2} -{\bm v}^T{\bm H}^{-1}{\bm v}\Bigg)\Delta^4
\nonumber\\ 
\end{eqnarray}
with Eqs.~(\ref{GapEqDelta0})~and~(\ref{GammaExp}), where all derivatives are evaluated at
$(\Delta^2,\boldsymbol{\Phi})=(0,\boldsymbol{\Phi}_0)$. Thus, 
we can get the GL coefficients for $\Delta^2$ as
\begin{equation}
\begin{aligned}
    c &=
   2 \left.
    \frac{\partial^2\Omega}{\partial\Delta^2}
    \right|_{\D=0,\boldsymbol{\F}=\boldsymbol{\F}_0},
    \\
    b &=2
    \left.
    \left(
    \frac{\partial^2\Omega}{\partial(\Delta^2)^2}
    -
    \boldsymbol{v}^{T}{\bm H}^{-1}\boldsymbol{v}
    \right)
    \right|_{\D=0,\boldsymbol{\F}=\boldsymbol{\F}_0}.\label{eq:bfull}
\end{aligned}
\end{equation}
The contribution ${\bm v}^T{\bm H}^{-1}{\bm v}$ accounts for the corrections from implicit $\Delta^2$ dependencies of ${\bm \Phi}$ as readily understood from Eqs.~(\ref{Hessian}) and~(\ref{vHv}).
For a stable normal phase the Hessian matrix ${\bm H}$ must be positive
definite, and accordingly ${\bm H}^{-1}$ is positive-semidefinite:
\begin{equation}
    \boldsymbol{v}^{T}{\bm H}^{-1}\boldsymbol{v} \geq 0\ . \label{vHv0}
\end{equation}
This property implies that the ${\bm \Phi}$ contributions always give a
negative correction to the coefficient of $\Delta^4$. In other words, multiple order parameters ${\bm \Phi}$ tend to reduce the effective quartic coefficient in such a way as to facilitate the first-order phase transition with respect to $\Delta$.

Let us return to the full thermodynamic potential~(\ref{eq:thermal potential}) in the main text with no renormalization applied. This potential can be divided into three parts:
\begin{eqnarray}
\O= V_{\rm MF} + V_q^{T=0} + V_q^{T\neq 0}\ , \label{OmegaFull}
\label{eq:potential form}
\end{eqnarray}
with $V_{\rm MF}$ being the mean-field potential defined in Eq.~(\ref{MFPotential}), while $V_q^{T=0}$ and $V_q^{T\neq0}$ are the $T$-independent and $T$-dependent quark one-loop contributions that read
\begin{eqnarray}
V_q^{T=0} &=& -2N_c \sum_{\xi={\rm p,a}}\int\frac{d^3k}{(2\p)^3}\e_{\xi} \ , \nonumber\\
V_q^{T\neq0} &=& -4N_cT\sum_{\xi={\rm p,a}}\int\frac{d^3k}{(2\p)^3}\ln(1+\me^{-\b\e_{\xi}}) \ . \nonumber\\
\end{eqnarray}
In the following analysis we will ignore the corrections from $M_l$ and $M_s$ corresponding to the ${\bm v}^T{\bm H}^{-1}{\bm v}$ in Eq.\eqref{eq:bfull}, so accordingly the coefficient of $\D^4$ can also be split into three parts
\begin{equation}
b = b_{\rm MF}+b_q^{T=0}+b_q^{T\ne0}\ 
\end{equation}
with
\begin{equation}
\begin{aligned}
b^{T=0}_q
&=\frac{1}{8}\sum_{\xi={\rm p,a}}\int_{\vec{k}}
\frac{1}{|E_l-\w_{\xi}\mu_I|^3},\\
b^{T\ne0}_q
&=-\frac{1}{4}\sum_{\xi={\rm p,a}}\int_{\vec{k}}
\begin{aligned}[t]
\Biggl[
&\frac{f(|E_l-\w_{\xi}\mu_I|)}
{|E_l-\w_{\xi}\mu_I|^3}
\\
+&\frac{f(|E_l-\w_{\xi}\mu_I|)\left(1-f(|E_l-\w_{\xi}\mu_I|)\right)}
{T|E_l-\w_{\xi}\mu_I|^2}
\Biggr] \, .
\end{aligned}
\end{aligned}
\end{equation}
One can confirm that $b_{\rm MF}>0$ and $b_q^{T=0}\geq0$ while $b_q^{T\ne0}<0$ when appropriate regularization schemes are adopted. Here, owing to the Boltzmann suppression $b_q^{T\neq0}$ is always suppressed as long as we stick comparably small $T$, for which the potential is mainly determined by $T$-independent $b_{\rm MF}$ and $b_q^{T=0}$. As we increase $T$, the negative $b_q^{T\ne0}$ dominates over the other two which results in a negative $b$. Thus, at higher temperatures we must take into account ${\cal O}(\Delta^6)$ terms particularly from $V_q^{T\neq0}$, which may yield the first-oder phase transition. This argument is independent of the regularization scheme, and therefore, one can expect that the first-order phase transition always occurs within a mean-field analysis at quark one-loop, when the phase-transition temperature becomes adequately high~\cite{Ferreira:2025zeu,Kamikado:2012bt}. The corrections from the implicit $\Delta^2$ dependencies of $M_l$ and $M_s$ further make it easier to induce the first-order transition. Different regularization schemes only affect the quantitative values of the critical temperatures for the first- and second-order phase transitions.

\begin{figure}[htbp]
\includegraphics[width=1\linewidth]{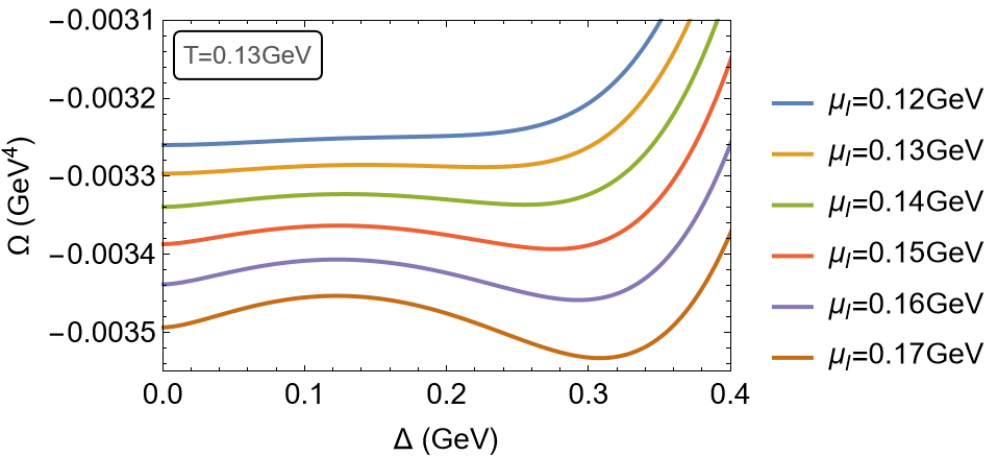}
\caption{The effective potential $\Omega$ as a function of $\Delta$ at $T=0.13\, \text{GeV}$, with different $\mu_I$.}
\label{fig:potential}
\end{figure}

Depicted in Fig.~\ref{fig:potential} denotes the $\Delta$ dependence of the effective potential $\Omega$ in Eq.~(\ref{OmegaFull}) at $T=0.13$ GeV with several $\mu_I$, where no expansions are done. In plotting this figure, we have employed the no-sea approximation as done in the main text with the parameters in Table~\ref{tab:Parameter}. When $\m_I$ is small $\D=0$ is the global minimum. Meanwhile, as the $\m_I$ is increased, the other minimum with finite $\Delta$ evolves and eventually becomes the global minimum, which signals the first-order phase transition.

Finally, we discuss the coefficient of $\Delta^2$:
\begin{equation}
\frac{c}{2}=c_{\rm MF}+c_q^{T=0}+c_q^{T\ne0}\ .
\end{equation}
From the above expressions, one can easily confirm that the thermal contribution $c_q^{T\ne0}$ is positive and  increases with temperature. Since the onset of pion condensation is determined by the condition $c=0$, a higher temperature requires a larger critical isospin chemical potential to trigger the phase transition. On the other hand, if the overall coefficient $c$ remained positive for all $\mu_I$, the thermodynamic potential would always have a unique minimum at $\Delta=0$, and spontaneous symmetry breaking would never occur. In our model, however, the mesonic kinetic term contributes a negative term $-2\mu_I^2\Delta^2/g^2$ to $c$, which guarantees that the overall coefficient can become negative at sufficiently large $\mu_I$, irrespective of the temperature. Therefore, in our calculation, the pion condensate strength will keep increasing with isospin chemical potential. But in NJL model, the strength will fall down if isospin chemical potential is large enough\cite{He:2005nk,Zhang:2025stm}.

\section{$\pi_0$ mass in the superfluid phase}
\label{ap:PionMass}

The $\pi_0$ mass in the pion superfluid phase is analytically determined to be
\begin{eqnarray}
M_{\pi_0} = 2\mu_I\ , \label{Pi0MassLinear}
\end{eqnarray}
as mentioned in Sec.~\ref{sec:meson}. Here we prove this noteworthy property.

The $\pi_0$ mass, $M_{\pi_0}$, is evaluated at a pole position of the propagator $D_{\pi_3}(p_0,\vec{p})$ with respect to $p_0$ at the rest frame. Hence, provided that $M_{\pi_0}=2\mu_I$, the following condition holds:
\begin{eqnarray}
D_{\pi_3}^{-1}(2\mu_I,\vec{0})  = (2\mu_I)^2-m_{\pi_3\pi_3}^2-\Pi_{\pi_3\pi_3}(2\mu_I,\vec{0}) = 0\ , \nonumber\\ \label{PolePi0}
\end{eqnarray}
where the self-energy part is of the form
\begin{eqnarray}
\Pi_{\pi_3\pi_3}(2\mu_I,\vec{0}) =  4g^2N_c\sum_{\xi={\rm p,a}}\int_{\vec{k}}\frac{f(\e_\xi)}{\e_\xi}
\end{eqnarray}
from Eq.~(\ref{Pi3Pi3}), with the help of an identity ($\zeta=\pm$)
\begin{eqnarray}
1-\zeta\frac{E_l^2-\mu_I^2+\Delta^2}{\epsilon_{\rm p}\epsilon_{\rm a}} =\zeta \frac{(2\mu_I)^2-(\epsilon_{\rm p}-\zeta\epsilon_{\rm a})^2}{2\epsilon_{\rm p}\epsilon_{\rm a}}\ .
\end{eqnarray}

Meanwhile, from the gap equation for $\Delta$, Eq.~(\ref{eq:gap equation}), it can be easily checked the nontrivial solution must satisfy
\begin{eqnarray}
4\mu_I^2 -m_{\pi_3\pi_3}^2 - \Pi_{\pi_3\pi_3}(2\mu_I,\vec{0})=0\ ,
\end{eqnarray}
with the tree-level mass~(\ref{eq:pion3 tree mass}). This equation is identical to the pole condition~(\ref{PolePi0}). Therefore, in the pion superfluidity $\pi_0$ mass is analytically determined by Eq.~(\ref{Pi0MassLinear}), regardless of temperatures.

\section{Discussion on the no-sea approximation}
\label{ap:no-sea}

In this work, we have adopted the no-sea approximation to evaluate the quark one-loop contributions to the thermodynamic potential. This treatment does not correspond to a conventional renormalization scheme, as it amounts to neglecting the vacuum fluctuations of the Dirac sea and retaining only the contributions from thermally excited quasiparticles. Given the scope of this study, we briefly examine the applicability and limitations of this approximation.

From the perspective of low-energy effective field theories, the no-sea approximation can be regarded as a phenomenological choice. Models such as the quark-meson model are commonly employed to investigate qualitative features of QCD and the associated patterns of symmetry breaking and restoration~\cite{Lenaghan:2000ey,Schaefer:2008hk,Mao:2009aq,Mintz:2012mz}. In the present framework, the model parameters are calibrated to reproduce selected vacuum observables, including experimental meson masses and decay constants, thereby incorporating phenomenological information about the vacuum into the model parameters. While the no-sea approximation omits fermion vacuum fluctuations, it provides a relatively simple framework for focusing on thermal medium contributions and studying the resulting response and phase structure of the system, without introducing the additional technical complications associated with the explicit treatment of vacuum fluctuations.

\begin{figure}[htbp]
\includegraphics[width=1\linewidth]{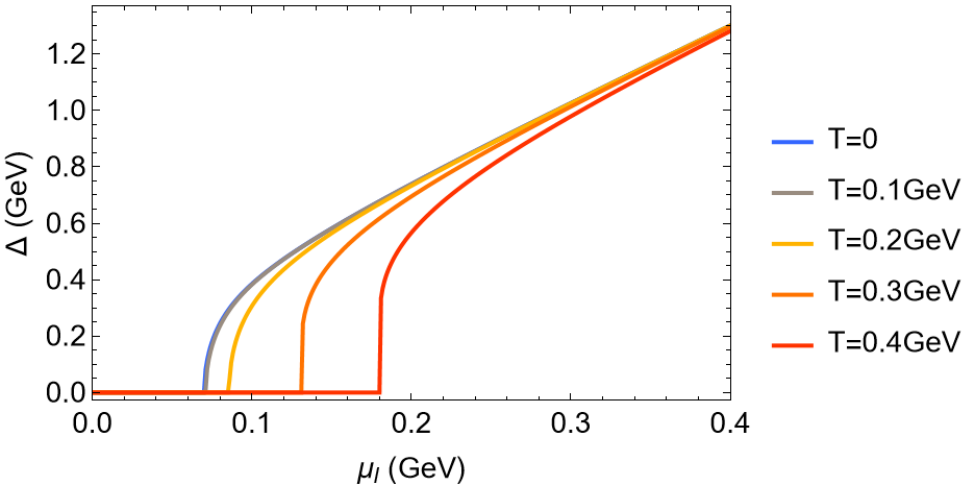}
\caption{The pion condensate $\D$ as a function of $\m_I$ at different temperatures, using the $\overline{\rm MS}$ scheme.}
\label{fig:pion condensate msbar}
\end{figure}

To examine the sensitivity of our results to the treatment of fermion vacuum fluctuations, we have also recalculated the pion condensate using the $\overline{\text{MS}}$ renormalization scheme. The results are shown in Fig.~\ref{fig:pion condensate msbar}. As seen from the figure, the pion condensate $\Delta$ exhibits qualitatively similar behavior as a function of the isospin chemical potential $\mu_I$ to that obtained in the no-sea approximation in Sec.~\ref{sec:phase}. In particular, the transition remains continuous at lower temperatures, while a discontinuous jump occurs at higher temperatures. This suggests that the first-order transition observed in our mean-field calculation is not simply a consequence of neglecting the fermion vacuum fluctuations. In addition, at large $\mu_I$, the condensate exhibits an approximately linear dependence on $\mu_I$ over the temperature range considered. These qualitative similarities indicate that the main features of the pion-condensation phase structure discussed in this work are not strongly sensitive to the treatment of the fermion vacuum contribution.

However, the no-sea approximation inherently omits fermion vacuum fluctuations, which can lead to quantitative limitations. In particular, the absence of vacuum contributions to the thermodynamic potential can affect the quantitative location of phase boundaries and the values of thermodynamic observables. Consequently, while the approximation provides a simple framework for studying thermal medium effects and qualitative phase structures, it should be regarded as a simplified treatment rather than a quantitatively complete description of the system. A full treatment of the fermion vacuum fluctuations would therefore be desirable for more quantitative studies.

\begin{figure}[htbp]
\centering
\includegraphics[width=0.9\linewidth]{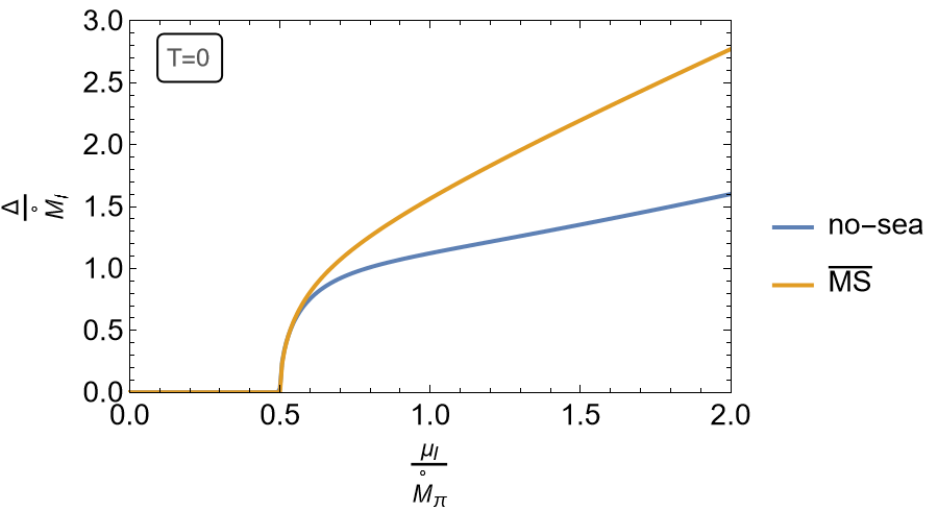}
\vspace{0.4cm}
\includegraphics[width=0.9\linewidth]{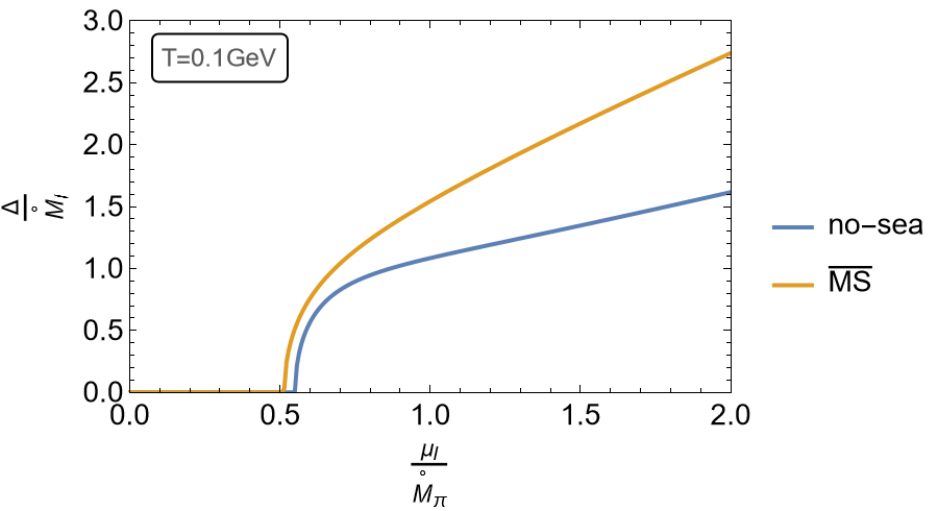}
\caption{$\mu_I$ dependencies of the pion condensate scaled by the vacuum pion mass, at $T=0$ (top) and $T=0.1$ GeV (bottom), within the two regularization scheme.}
\label{fig:ns vs msbar}
\end{figure}

To further illustrate the quantitative differences between the two treatments, a direct comparison of the pion condensate $\Delta$ as a function of the isospin chemical potential at $T = 0$ and $T = 0.1\text{ GeV}$ is shown in Fig.~\ref{fig:ns vs msbar}. The two calculations give rather similar results in the vicinity of the critical point, while noticeable quantitative deviations develop as $\mu_I$ increases. More generally, the differences between the two treatments can become more pronounced under thermodynamic conditions where vacuum fluctuations have a stronger impact on the phase structure. These comparisons suggest that the no-sea approximation is useful for capturing the qualitative phase structure and symmetry patterns considered in this work, but that the inclusion of fermion vacuum fluctuations becomes increasingly important for quantitative predictions away from the vicinity of the critical region.

\bibliographystyle{apsrev4-1}
\bibliography{paper_v4}

\end{document}